%% file: p90r.tex
\documentclass[11pt]{article}
\usepackage{amsmath}
\usepackage{amsfonts}
\usepackage{amssymb}
\usepackage{graphicx}

\def\le{\leq}
\def\ge{\geq}

\newcommand{\ls}[1]
   {\dimen0=\fontdimen6\the\font \lineskip=#1\dimen0
\advance\lineskip.5\fontdimen5\the\font \advance\lineskip-\dimen0
\lineskiplimit=.9\lineskip \baselineskip=\lineskip
\advance\baselineskip\dimen0 \normallineskip\lineskip
\normallineskiplimit\lineskiplimit \normalbaselineskip\baselineskip
\ignorespaces }

\newtheorem{theorem}{Theorem}

\newcommand {\dfn} {\stackrel{\Delta} {=}}

\newcommand {\reals} {{\rm I\!R}}

\newcommand {\hU} {\hat{U}}
\newcommand {\hu} {\hat{u}}

\newcommand{\calA}{{\cal A}}
\newcommand{\calB}{{\cal B}}
\newcommand{\calC}{{\cal C}}

\newcommand{\calK}{{\cal K}}

\newcommand{\calP}{{\cal P}}

\newcommand{\calU}{{\cal U}}
\newcommand{\calV}{{\cal V}}

\newcommand{\calX}{{\cal X}}
\newcommand{\calY}{{\cal Y}}
\newcommand{\calZ}{{\cal Z}}

\begin{document}
\thispagestyle{empty}
\title{On Joint Coding for Watermarking and Encryption}
\author{Neri Merhav}
\date{}
\maketitle

\begin{center}
Department of Electrical Engineering \\
Technion - Israel Institute of Technology \\
Haifa 32000, ISRAEL \\
{\tt merhav@ee.technion.ac.il}
\end{center}
\vspace{1.5\baselineskip}
\setlength{\baselineskip}{1.5\baselineskip}

\begin{abstract}
In continuation to earlier works where the problem of joint
information embedding and lossless compression (of the composite signal)
was studied in the absence \cite{MM03} and in the presence \cite{MM04}
of attacks, here we consider the additional ingredient of
protecting the secrecy of the watermark against an unauthorized party, which
has no access to a secret key shared by the legitimate parties.
In other words, we study the problem of joint
coding for three objectives: information embedding, compression, and encryption.
Our main result is a coding theorem that provides a 
single--letter characterization of the best achievable tradeoffs among
the following parameters: the distortion between the composite signal and
the covertext, the distortion in reconstructing the watermark by the legitimate
receiver, the compressibility of the composite signal (with and without the key),
and the equivocation of the watermark, as well as its reconstructed
version, given the composite signal. In the attack--free case, if the key
is independent of the covertext, this coding 
theorem gives rise to a {\it threefold} separation 
principle that tells that asymptotically, for long block codes, no optimality
is lost by first applying a rate--distortion code to the watermark source,
then encrypting the compressed codeword, and finally, embedding it into the
covertext using the embedding scheme of \cite{MM03}. In the more general case,
however, this separation principle is no longer valid, as the key plays an
additional role of side information used by the embedding unit.
 
\vspace{1cm}

\end{abstract}

\newpage
\section{Introduction}

It is common to say that encryption and watermarking (or information hiding)
are related but they are substantially 
different in the sense that in the former,
the goal is to protect the secrecy of 
the {\it contents} of information, whereas in
the latter, it is the very {\it existence} 
of this information that is to be kept
secret. 

In the last few years, however, we are witnessing increasing 
efforts around the {\it combination} of encryption and watermarking,
which is motivated by the desire to 
further enhance the security of sensitive information that is
being hidden in the host signal. This is to guarantee that even 
if the watermark is somehow detected by a hostile
party, its contents still remain secure due to the encryption.
This combination of watermarking and encryption 
can be seen both in recently reported research work (see, e.g., 
\cite{AKS02},\cite{CC02},\cite{JML00},\cite{KNSTN02},\cite{MW04},\cite{SIA99} 
and references therein)
and in actual technologies used in 
commercial products with a copyright protection 
framework, such as the CD and the DVD. Also, some commercial companies 
that provide Internet documents, have in their
websites links to copyright warning messages, 
saying that their data are protected by
digitally encrypted watermarks (see, e.g., 
{\tt http://genealogy.lv/1864Lancaster/copyright.htm}).

This paper is devoted to the information--theoretic
aspects of joint watermarking and encryption together with 
lossless compression of the composite signal that 
contains the encrypted watermark. Specifically, we extend
the framework studied in \cite{MM03} and \cite{MM04}
of joint watermarking and compression, so as to include
encryption using a secret key. Before we describe the
setting of this paper concretely, we pause then to give some
more detailed background on the work reported in \cite{MM03} and 
\cite{MM04}.

In \cite{MM03}, the following problem was studied: Given a 
covertext source vector $X^n=(X_1,\ldots,X_n)$, generated by
a discrete memoryless source (DMS), and a message $m$, 
uniformly distributed in $\{1,2,\ldots, 2^{nR_e}\}$, independently of $X^n$,
with $R_e$ designating the embedding rate, we wish to 
generate a composite (stegotext) vector
$Y^n=(Y_1,\ldots,Y_n)$ that satisfies the 
following requirements: (i) Similarity to the
covertext (for reasons of maintaining quality), in the sense that a distortion 
constraint, $Ed(X^n,Y^n)=\sum_{t=1}^nEd(X_t,Y_t)\le nD$,
holds, (ii) compressibility (for reasons of saving storage space and bandwidth), in the sense 
that the normalized entropy, $H(Y^n)/n$, does not exceed some
threshold $R_c$, and (iii) reliability in decoding the message $m$ 
from $Y^n$, in the sense that
the decoding error probability is arbitrarily 
small for large $n$. A single--letter characterization
of the best achievable tradeoffs among $R_c$, $R_e$, 
and $D$ was given in \cite{MM03}, and was
shown to be achievable by an extension 
of the ordinary lossy source coding theorem, giving rise to the
existence of $2^{nR_e}$ {\it disjoint} 
rate--distortion codebooks (one per each possible watermark
message) as long as $R_e$ does not exceed
a certain fundamental limit. In \cite{MM04}, this setup 
was extended to include 
a given memoryless attack channel, 
$P(Z^n|Y^n)$, where item (iii) above was redefined such that
the decoding was based on $Z^n$ rather than on $Y^n$, and where, in view of requirement (ii),
it is understood that the attacker has access to the compressed version of $Y^n$,
and so, the attacker decompresses $Y^n$ before the attack and re--compresses it after.
This extension from [8] to [9] involved
an different approach, which was in the 
spirit of the Gel'fand--Pinsker coding theorem for
a channel with non--causal side information 
(SI) at the transmitter \cite{GP80}. The role of SI, in this case, was
played by the covertext.

In this paper, we extend the settings of \cite{MM03} and \cite{MM04}
to include encryption. For the sake of clarity of the exposition,
we do that in several steps.

\begin{figure}[h]
\hspace*{-2cm}\input{p90fig1.pstex_t}
\caption{A generic watermarking/encryption system.}
\label{gen}
\end{figure}
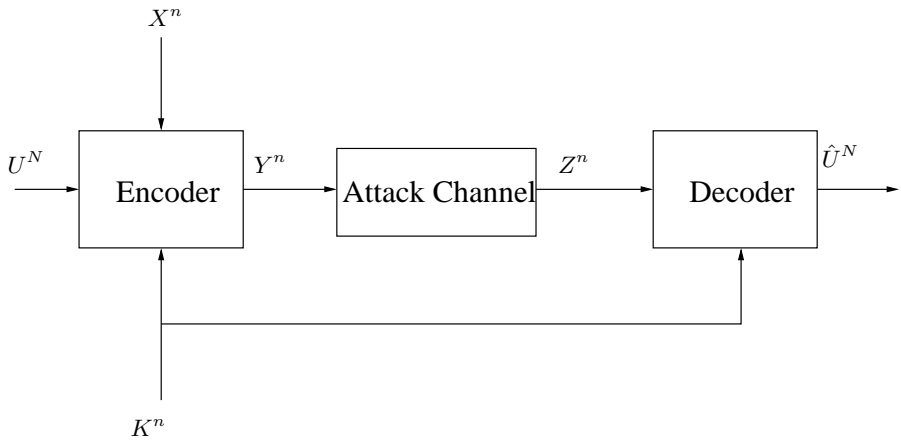

In the first step, we extend the attack--free setting of 
\cite{MM03}: In addition to including encryption,
we also extend the model of the watermark message
source to be an arbitrary DMS, $U_1,U_2,\ldots$, 
independent of the covertext, 
and not necessarily a binary symmetric source 
(BSS) as in \cite{MM03} and \cite{MM04}.
Specifically, we now assume that the encoder
has three inputs (see Fig.\ \ref{gen}): The covertext source vector, 
$X^n$, an independent (watermark) message source vector 
$U^N=(U_1,\ldots,U_N)$, where $N$ may differ from 
$n$ if the two sources operate
in different rates, and a secret key (shared also with 
the legitimate decoder) $K^n=(K_1,\ldots,K_n)$, which, for mathematical
convenience, is assumed to operate at the same 
rate as the covertext. It is assumed, 
at this stage, that $K^n$
is independent of $U^N$ and $X^n$.
Now, in addition to requirements
(i)-(iii), we impose a requirement on the equivocation 
of the message source relative to an
eavesdropper that has access to $Y^n$, but not
to $K^n$. Specifically, we would like the normalized
conditional entropy, $H(U^N|Y^n)/N$, to exceed
a prescribed threshold, $h$ (e.g., $h=H(U)$ 
for perfect secrecy). Our first result is 
a coding theorem that gives a set of necessary
and sufficient conditions, in terms of single--letter inequalities,
such that a triple $(D,R_c,h)$ is achievable,
while maintaining 
reliable reconstruction of $U^N$ at the legitimate receiver.

In the second step, we relax the requirement of perfect (or almost
perfect) watermark reconstruction, and assume that we are willing to
tolerate a certain distortion between the watermark message $U^N$ and its
reconstructed version $\hat{U}^N$, that is,
$Ed'(U^N,\hat{U}^N)=\sum_{i=1}^NEd'(U_i,\hat{U}_i)\le ND'$. For example,
if $d'$ is the Hamming distortion measure then $D'$, of course, designates 
the maximum allowable bit error probability (as opposed to the block error
probability requirement of \cite{MM03} and \cite{MM04}). Also, in this case,
it makes sense to impose a requirement regarding the
equivocation of the {\it reconstructed} message,
$\hat{U}^N$, namely, $H(\hat{U}^N|Y^n)/N\ge h'$, for some prescribed
constant $h'$. The rationale is that it is $\hat{U}^N$, 
not $U^N$, that
is actually conveyed to the legitimate receiver, and hence
there is an incentive to protect the secrecy of $\hat{U}^N$.
We will take into account both
equivocation requirements, with the understanding that if one of them
is superfluous, then the corresponding threshold 
($h$ or $h'$ accordingly) can always be set to zero.
Our second result then extends the above--mentioned
coding theorem to a single--letter characterization of achievable quintuples
$(D,D',R_c,h,h')$. As will be seen, 
this coding theorem gives rise
to a threefold separation theorem, that separates, without asymptotic loss
of optimality, between three stages: rate--distortion coding of $U^N$,
encryption of the compressed bitstream, and finally, embedding the
resulting encrypted version using the embedding scheme of \cite{MM03}.
The necessary and sufficient conditions related to the
encryption are completely decoupled from those of the 
embedding and the stegotext compression.

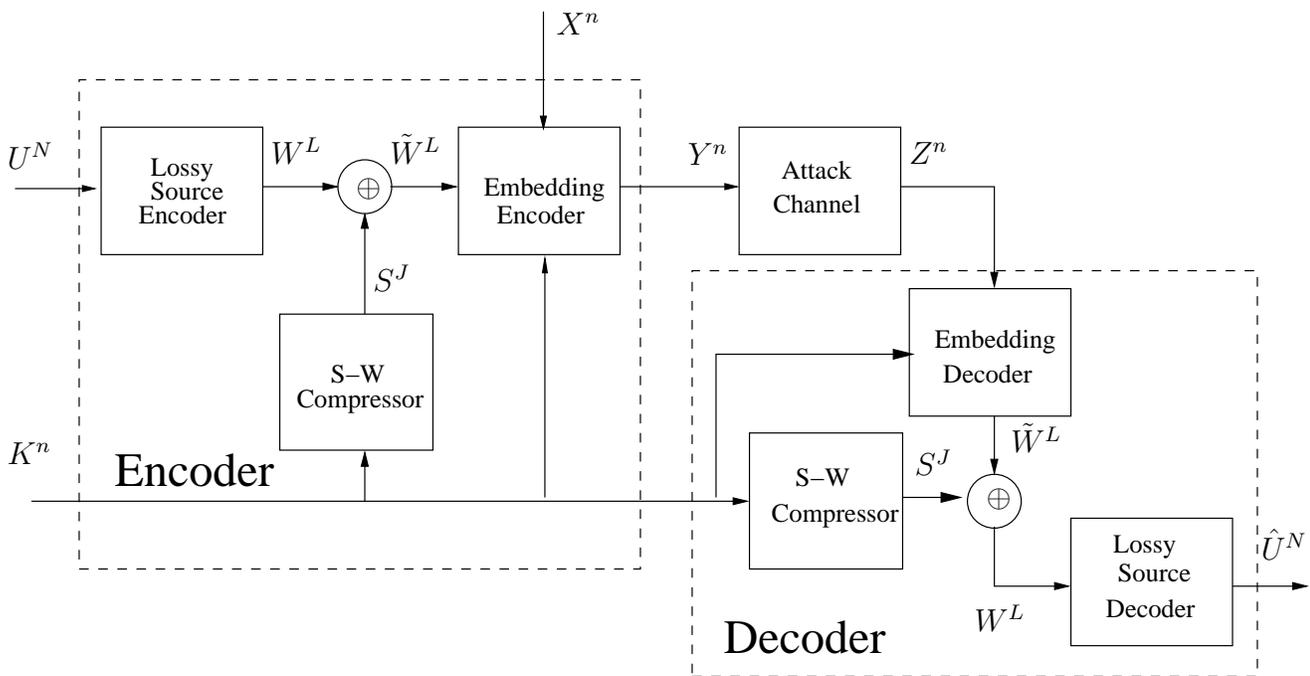
\begin{figure}[h]
\hspace*{-2cm}\input{p90fig2.pstex_t}
\caption{The proposed watermarking/encryption scheme (general case).}
\label{dir}
\end{figure}

In the third and last step, we drop the assumption of an attack--free
system and we assume a given memoryless attack channel, in analogy to \cite{MM04}.
Again, referring to Fig.\ \ref{gen}, it should be understood that the stegotext $Y^n$ is
stored (or transmitted) in compressed form, and that the attacker decompresses $Y^n$
before the attack and decompresses after (the compression and decompression units
are omitted from the figure).
As it will turn out, in the case of a memoryless attack, there is an interaction between the
encryption and the embedding, even if the key is still
assumed independent of the covertext. In particular,
it will be interesting to see 
that the key, in addition to its original role in encryption,
serves as SI that is available
to both encoder and decoder (see Fig.\ \ref{dir}).
Also, because of the dependence between
the key and the composite signal, and the fact that the key is available
to the legitimate decoder as well,
it is reasonable 
to let the compressibility constraint correspond also to the
conditional entropy of $Y^n$ given $K^n$, that is, {\it private} compression as opposed
to the previously considered
{\it public} compression, without the key, which enables decompression but not decryption
(when these two operations are carried out by different, remote units).
Accordingly, we will consider both the conditional and the unconditional
entropies of $Y^n$, i.e.,
$H(Y^n)/n\le R_c$ and $H(Y^n|K^n)/n\le R_c'$.
Our final result then is a coding theorem that 
provides a single--letter characterization of the
region of achievable six--tuples $(D,D',R_c,R_c',h,h')$. 
Interestingly, this characterization remains essentially
unaltered even if there is dependence between the key and the covertext,
which is a reasonable thing 
to have once the key and the stegotext interact in the first place.\footnote{In fact,
the choice of the conditional distribution $P(K^n|X^n)$ is a degree of freedom
that can be optimized subject to the given randomness resources.}
In this context, the system designer confronts an interesting
dilemma regarding the desirable degree of statistical dependence
between the key and the covertext, which affects the dependence
between the key and the stegotext.
On the one hand, strong dependence can reduce
the entropy of $Y^n$ given $K^n$ (and thereby reduce 
$R_c'$), and can also help in the embedding 
process: For example, the extreme case of $K^n=X^n$ (which
corresponds to {\it private} watermarking since
the decoder actually has access to the covertext) is particularly
interesting because in this case, 
for the encryption key, there is no need for any external resources
of randomness, in addition to 
the randomness of the covertext that is already available. 
On the other hand, when there is strong dependence between $K^n$ and $Y^n$,
the secrecy of the watermark might be sacrificed since $H(K^n|Y^n)$
decreases as well. 
An interesting point, in this context, is that the
Slepian--Wolf encoder \cite{SW73} (see Fig.\ \ref{dir}) is used to 
generate, from $K^n$, random bits that are essentially 
independent of $Y^n$
(as $Y^n$ is generated only after the encryption).
These aspects will be seen in detail in Section 4, and even more so, in Section 6.

The remaining parts of this paper are organized as follows:
In Section 2, we set some notation conventions. 
Section 3 will be devoted to a formal problem description
and to the presentation of the main result for the attack--free case
with distortion--free watermark reconstruction (first step
described above). In Section 4, the setup and the results will
be extended along the lines of the second and the third steps,
detailed above,
i.e., a given distortion level in the watermark reconstruction and the
incorporation of an attack channel. Finally, Sections 5 and 6 
will be devoted to the proof of the last (and most general) version of the
coding theorem, with Section 5 focusing on the converse part,
and Section 6 -- on the direct part. 

\section{Notation Conventions} 

We begin by establishing some notation conventions.
Throughout this paper, scalar random 
variables (RV's) will be denoted by capital
letters, their sample values will be denoted by
the respective lower case letters, and their alphabets will be denoted
by the respective calligraphic letters.
A similar convention will apply to
random vectors and their sample values,
which will be denoted with same symbols superscripted by the dimension.
Thus, for example, $A^\ell$ ($\ell$ -- positive integer)
will denote a random $\ell$-vector $(A_1,...,A_\ell)$,
and $a^\ell=(a_1,...,a_\ell)$ is a specific vector value in $\calA^\ell$,
the $\ell$-th Cartesian power of $\calA$. The 
notations $a_i^j$ and $A_i^j$, where $i$
and $j$ are integers and $i\le j$, will designate segments $(a_i,\ldots,a_j)$
and $(A_i,\ldots,A_j)$, respectively,
where for $i=1$, the subscript will be omitted (as above). 
For $i > j$, $a_i^j$ (or $A_i^j$) will be understood as the null string.
Sequences without specifying indices are denoted by $ \{\cdot\} $.

Sources and channels will be denoted generically by the letter $P$, or $Q$,
subscripted by the name of the RV and its conditioning,
if applicable, e.g., $P_U(u)$ is the probability function of
$U$ at the point $U=u$, $P_{K|X}(k|x)$
is the conditional probability of $K=k$ given $X=x$, and so on.
Whenever clear from the context, these subscripts will be omitted.
Information theoretic quantities like entropies and mutual
informations will be denoted following the usual conventions
of the information theory literature, e.g., $H(U^N)$, $I(X^n;Y^n)$,
and so on. For single--letter
information quantities (i.e., when $n=1$ or $N=1$), 
subscripts will be omitted, e.g., $H(U^1)=H(U_1)$ will
be denoted by $H(U)$, 
similarly, $I(X^1;Y^1)=I(X_1;Y_1)$ will be denoted by $I(X;Y)$, and so on.

\section{Problem Definition and Main Result for Step 1}

We now turn to the formal description 
of the model and the problem setting for step 1,
as described in the Introduction.
A source $P_X$, henceforth referred to as the 
{\it covertext source} or the {\it host source}, generates a sequence of
independent copies, $\{X_t\}_{t=-\infty}^{\infty}$, of a finite--alphabet RV,
$X\in\calX$. At the same time and independently,
another source $P_U$, henceforth referred to as the {\it message source}, 
or the {\it watermark source}, generates a sequence of
independent copies, $\{U_i\}_{i=-\infty}^{\infty}$, of a finite--alphabet RV,
$U\in\calU$. The relative rate between the message source and the covertext
source is $\lambda$ message symbols per covertext symbol. This means that
while the covertext 
source generates a block of $n$ symbols, say, $X^n=(X_1,\ldots,X_n)$,
the message source generates a block 
of $N=\lambda n$ symbols, $U^N=(U_1,\ldots,U_N)$
(assuming, without essential loss of 
generality, that $\lambda n$ is a positive integer).
In addition to the covertext source and 
the message source, yet another source, $P_K$,
henceforth referred to as the {\it key source},
generates a sequence of independent copies, 
$\{K_t\}_{t=-\infty}^{\infty}$, of a finite--alphabet RV,
$K\in\calK$, independently\footnote{The assumption 
of independence between $\{K_t\}$
and $\{X_t\}$ is temporary and made now primarily for 
the sake of simplicity of the exposition. It will
be dropped later on.}
of both $\{X_t\}$ and $\{U_i\}$. 
The key source is assumed to operate at the same
rate as the covertext source, that is, while 
the covertext source generates the block $X^n$
of length $n$,
the key source generates a block of $n$
symbols as well, $K^n=(K_1,\ldots,K_n)$.

Given $n$ and $\lambda$, a block code 
for {\it joint watermarking, encryption, and compression}
is a mapping $f_n:\calU^N\times\calX^n\times\calK^n\to\calY^n$, 
$N=\lambda n$, whose output
$y^n=(y_1,\ldots,y_n)=f_n(u^N,x^n,k^n)\in\calY^n$ 
is referred to as the {\it stegotext} or the
{\it composite signal}, and accordingly, the finite alphabet $\calY$ is
referred to as the {\it stegotext alphabet}. 
Let $d:\calX\times\calY\to \reals^+$ denote
a single--letter distortion measure between 
covertext symbols and stegotext symbols,
and let the distortion between the vectors, $x^n\in\calX^n$ and $y^n\in\calY^n$,
be defined additively across the corresponding components, as usual. 

An $(n,\lambda,D,R_c,h,\delta)$ code is a block code 
for joint watermarking, encryption, and compression,
with parameters $n$ and $\lambda$, that satisfies the following requirements:
\begin{itemize}
\item [1.] The expected distortion 
between the covertext and the stegotext satisfies
\begin{equation}
\sum_{t=1}^n Ed(X_t,Y_t)\le nD.
\end{equation}
\item [2.] The entropy of the stegotext
satisfies
\begin{equation}
H(Y^n)\le nR_c.
\end{equation}
\item [3.] The equivocation of the message source satisfies
\begin{equation}
H(U^N|Y^n)\ge Nh.
\end{equation}
\item [4.] There exists a
decoder $g_n:\calY^n\times\calK^n\to\calU^N$ such that
\begin{equation}
P_e\dfn\mbox{Pr}\{g_n(Y^n,K^n)\ne U^N\}\le \delta.
\end{equation}
\end{itemize}
For a given $\lambda$, a triple $(D,R_c,h)$ is said to be {\it achievable}
if for every $\epsilon > 0$, there is a sufficiently large $n$
for which 
$(n,\lambda,D+\epsilon,R_c+\epsilon,h-\epsilon,\epsilon)$ codes exist.
The {\it achievable region} of triples 
$(D,R_c,h)$ is the set of all achievable
triples $(D,R_c,h)$. For simplicity, it is assumed\footnote{At the end of Section 4 (after Theorem 4),
we discuss the case where this limitation (or its analogue in lossy reconstruction of $U^N$) is dropped.}
that $H(K)
\le\lambda H(U)$ as this upper limit on $H(K)$ suffices
to achieve perfect secrecy.

Our first coding theorem is the following:
\begin{theorem}
A triple $(D,R_c,h)$ is achievable if and only if the following conditions
are both satisfied:
\begin{itemize}
\item [(a)] $h \le H(K)/\lambda$.
\item [(b)] There exists a channel $\{P_{Y|X}(y|x),~x\in\calX,~y\in\calY\}$
such that: (i) $H(Y|X)\ge\lambda H(U)$, (ii) $R_c\ge\lambda H(U)+I(X;Y)$, and
(iii) $D\ge Ed(X,Y)$.
\end{itemize}
\end{theorem}

As can be seen, the encryption, on the one hand, and the embedding and the
compression, on the other hand, do not interact at all in this theorem.
There is a complete decoupling between them: While
condition (a) refers solely to the key and the secrecy of the watermark,
condition (b) is only about the embedding--compression 
part, and it is a replica of the conditions of 
the coding theorem in \cite{MM03}, where the role of the embedding rate, $R_e$
(see Introduction above),
is played by the product $\lambda H(U)$. This suggests a very simple separation 
principle, telling that in order to 
attain a given achievable triple $(D,R_c,h)$,
first compress the watermark $U^N$ to its entropy, then encrypt $Nh$ bits (out
of the $NH(U)$) of the compressed 
bit--string (by bit--by--bit XORing with the same number of
compressed key bits), and finally, embed this 
partially encrypted compressed bit--string into
the covertext, using the coding theorem of \cite{MM03} (again, 
see the Introduction
above for a brief description of this).

\section{Extensions to Steps 2 and 3}

Moving on to Step 2, we now relax requirement no.\ 4 in
the above definition of 
an $(n,\lambda,D,R_c,h,\delta)$ code, and allow a certain
distortion between $U^N$ and its reconstruction $\hat{U}^N$
at the legitimate decoder.
More precisely, let $\hat{\calU}$ denote a finite alphabet,
henceforth referred to as the {\it message reconstruction
alphabet}. Let $d':\calU\times\hat{\calU}\to\reals^+$ denote
a single--letter distortion measure between message symbols
and message reconstruction symbols, and let the distortion
between vectors $u^N\in\calU^N$ 
and $\hat{u}^N\in\hat{\calU}^N$ be again, defined additively
across the corresponding components. Finally, let $R_U(D')$
denote the rate--distortion function of the source $P_U$
w.r.t.\ $d'$, i.e.,
\begin{equation}
R_U(D')=\min\{I(U;\hU):~Ed'(U,\hU)\le D'\}.
\end{equation}
It will now be assumed that $H(K)\le\lambda R_U(D')$, for the
same reasoning as before.

Requirement no.\ 4 is now replaced by the
following requirement: 
There exists a
decoder $g_n:\calY^n\times\calK^n\to\hat{\calU}^N$ such that
$\hU^N=(\hU_1,\ldots,\hU_N)=g_n(Y^n,K^n)$ satisfies:
\begin{equation}
\label{distp}
\sum_{i=1}^NEd'(U_i,\hat{U}_i)\le ND'.
\end{equation}
In addition to this modification of requirement no.\ 4, we
add, to requirement no.\ 3, a specification regarding the
minimum allowed equivocation w.r.t.\ the reconstructed message:
\begin{equation}
H(\hU^N|Y^n)\ge Nh',
\end{equation}
in order to guarantee that the secrecy of the reconstructed message 
is also secure enough. Accordingly, we modify the above definition of a block
code as follows:
An $(n,\lambda,D,D',R_c,h,h')$ code is a block code 
for joint watermarking, encryption, and compression
with parameters $n$ and $\lambda$ that satisfies requirements 1--4,
with the above modifications of requirements 3 and 4.
For a given $\lambda$, a quintuple $(D,D',R_c,h,h')$ 
is said to be {\it achievable}
if for every $\epsilon > 0$, there is a sufficiently large $n$
for which 
$(n,\lambda,D+\epsilon,D'+\epsilon,R_c+\epsilon,h-\epsilon,h'-\epsilon)$ 
codes exist.

Our second theorem extends Theorem 1 to this setting:

\begin{theorem}
A quintuple $(D,D',R_c,h,h')$ is achievable 
if and only if the following conditions
are all satisfied:
\begin{itemize}
\item [(a)] $h\le H(K)/\lambda+H(U)-R_U(D')$.
\item [(b)] $h'\le H(K)/\lambda$.
\item [(c)] There exists a channel $\{P_{Y|X}(y|x),~x\in\calX,~y\in\calY\}$
such that: (i) $\lambda R_U(D')\le H(Y|X)$, 
(ii) $R_c\ge\lambda R_U(D')+I(X;Y)$, and
(iii) $D\ge Ed(X,Y)$.
\end{itemize}
\end{theorem}

As can be seen, the passage from Theorem 1 to Theorem 2 includes
the following modifications:
In condition (c), $H(U)$ is simply replaced by $R_U(D')$ as expected. 
This means that the lossless compression code of $U^N$, 
in the achievability of Theorem 1, is
now replaced by a rate--distortion code for distortion level $D'$.
Conditions (a) and (b) now tell us that the key rate (in terms of entropy)
should be sufficiently large to satisfy both equivocation requirements.
Note that the condition regarding the 
equivocation w.r.t.\ the clean message source is softer than
in Theorem 1 as $H(U)-R_U(D')\ge 0$. This is because the rate--distortion code
for $U^N$ already introduces an uncertainty of $H(U)-R_U(D')$ bits per
symbol, and so, the encryption 
should only complete it to the desired level of $h$
bits per symbol. This point is discussed in depth in \cite{Yamamoto97}.
Of course, by setting $D'=0$ (and hence also $h'=h$), we are back
to Theorem 1.

We also observe that the encryption and the embedding are still decoupled
in Theorem 2, and that an achievable quintuple can still be attained
by separation: First, apply a rate--distortion code to $U^N$, as mentioned
earlier, then encrypt $N\cdot\max\{h+R_U(D')-H(U),h'\}$ bits
of the compressed codeword (to satisfy both equivocation requirements),
and finally, embed the (partially) encrypted
codeword into $X^n$, again, by using the scheme of \cite{MM03}.
Note that without the encryption and without
requirement no.\ 2 of the compressibility of $Y^n$, this
separation principle is a special case of the one in
\cite{MS03}, where a separation theorem was established
for the Wyner--Ziv source (with SI correlated to the source at the decoder)
and the Gel'fand--Pinsker channel (with channel SI at the encoder).
Here, there is no SI correlated to the source
and the role of channel SI is fulfilled by the covertext.
Thus, the new observation here is that the separation theorem continues
to hold in the presence of encryption and requirement no.\ 2.

Finally, we turn to step 3, of including an attack channel (see Fig.\ 
\ref{gen}).
Let $\calZ$ be a finite alphabet, henceforth referred to as
the {\it forgery alphabet}, and 
let $\{P_{Z|Y}(z|y),~y\in\calY,~z\in\calZ\}$
denote a set of conditional PMF's from the stegotext alphabet to
the forgery alphabet. We now assume that the stegotext vector
is subjected to an attack modelled by the memoryless channel,
\begin{equation}
P_{Z^n|Y^n}(z^n|y^n)=\prod_{t=1}^n P_{Z|Y}(z_t|y_t).
\end{equation}
The output $Z^n$ of the attack channel will henceforth be referred to as
the {\it forgery}. 

It is now assumed 
and that the legitimate decoder has access to $Z^n$, 
rather than $Y^n$ (in addition, of course, to $K^n$).
Thus, in requirement no.\ 4, the
decoder is redefined again, this time,
as a mapping $g_n:\calZ^n\times\calK^n\to\hat{\calU}^N$ 
such that $\hat{U}^N=g_n(Z^n,K^n)$ satisfies
the distortion constraint (\ref{distp}). As for the equivocation requirements,
the conditioning will now be on both $Y^n$ and $Z^n$, i.e.,
\begin{equation}
H(U^N|Y^n,Z^n)\ge Nh~~~\mbox{and}~~~
H(\hU^N|Y^n,Z^n)\ge Nh',
\end{equation}
as if the
attacker and the eavesdropper are the same party (or if they cooperate),
then s/he may access both. In fact, 
for the equivocation of $U^N$, the conditioning on $Z^n$ is immaterial
since $U^N\to Y^n\to Z^n$ is always a Markov chain, but it is not clear
that $Z^n$ is superfluous for the equivocation w.r.t.\ $\hU^N$
since $Z^n$ is one of the inputs to the decoder whose output is $\hU^N$.
Nonetheless, for the sake of uniformity and convenience (in the proof),
we keep the conditioning on $Z^n$ in both equivocation criteria.

Redefining block codes and achievable quintuples $(D,D',R_C,h,h')$ 
according to the modified requirements in the same spirit, we now
have the following coding theorem, which is substantially
different from Theorems 1 and 2:

\begin{theorem}
A quintuple $(D,D',R_c,h,h')$ is achievable 
if and only if there exist RV's $V$ and $Y$ such that
$P_{KXVYZ}(k,x,v,y,z)=P_X(x)P_K(k)P_{VY|KX}(v,y|k,x)P_{Z|Y}(z|y)$,
where the alphabet size of $V$ is bounded by $|\calV|\le
|\calK|\cdot|\calX|\cdot|\calY|+1$, and such that the following
conditions are all satisfied:
\begin{itemize}
\item [(a)] $h\le H(K|Y)/\lambda+H(U)-R_U(D')$.
\item [(b)] $h'\le H(K|Y)/\lambda$.
\item [(c)] $\lambda R_U(D')\le I(V;Z|K)-I(V;X|K)$.
\item [(d)] $R_c\ge \lambda R_U(D')+I(X;Y,V|K)+I(K;Y)$.
\item [(e)] $D \ge Ed(X,Y)$.
\end{itemize}
\end{theorem}

First, observe that here, unlike in Theorems 1 and 2, it is no longer
true that the encryption and the embedding (along with stegotext compression) 
are decoupled, yet
the rate--distortion compression of $U^N$ is still separate and decoupled from both.
In other words, the separation principle applies here in partial manner only.
Note that now, although $K$ is still assumed independent of $X$,
it may, in general, depend on $Y$. On the negative side, 
this dependence causes a reduction
in the equivocation of both the message source and its reconstruction,
and therefore $H(K|Y)$ replaces $H(K)$ in conditions (a) and (b).
On the positive side, on the
other hand, this dependence introduces new degrees of freedom in
enhancing the tradeoffs between the embedding performance
(condition (c)) and the compressibility (condition (d)). 

The achievability of Theorem 3 involves essentially
the same stages as before (rate--distortion coding of $U^N$, followed by
encryption, followed in turn by embedding), but this time,
the embedding scheme is a conditional version of the one proposed in
\cite{MM04}, where all codebooks depend on $K^n$, the SI given at
both ends (see Fig.\ \ref{dir}). 
An interesting point regarding the encryption is that
one needs to generate, from $K^n$, essentially $nH(K|Y)$ random bits that
are {\it independent} of $Y^n$ (and $Z^n$), in order to protect the
secrecy against an eavesdropper that observes $Y^n$ and $Z^n$.
Clearly, if $Y^n$ was given in advance to the encrypting unit, then
the compressed bitstring of an optimal lossless
source code that compresses $K^n$, given $Y^n$ as SI, would have
this property (as if there was any dependence, then this bitstring could have
been further compressed, which is a contradiction). 
However, such a source code 
cannot be implemented
since $Y^n$ itself is generated 
from the encrypted message, i.e., {\it after}
the encryption. In other words, this would 
have required a circular mechanism, which may not be feasible.
A simple remedy is then to use a 
{\it Slepian--Wolf encoder} \cite{SW73}, that generates
$nH(K|Y)$ bits that are essentially 
independent of $Y^n$ (due to the same consideration), 
without the need to access the
vector $Y^n$ to be generated.
For more details, the reader is referred to the 
proof of the direct part (Section 6).

Observe that in the absence of attack (i.e., $Z=Y$),
Theorem 2 is obtained as a special case 
of Theorem 3 by choosing $V=Y$ and letting
both be independent of $K$, a choice which is simultaneously the best for
conditions (a)--(d) of Theorem 3. To see this, note the following simple
inequalities:
In conditions (a) and (b), $H(K|Y)\le H(K)$. In condition (c), 
by setting $Z=Y$, we have
\begin{eqnarray}
I(V;Y|K)-I(V;X|K)&\le&I(V;X,Y|K)-I(V;X|K)\nonumber\\
&=&I(V;Y|X,K)\nonumber\\
&\le&H(Y|X,K)\nonumber\\
&\le&H(Y|X).
\end{eqnarray}
Finally in condition (d), clearly, $I(K;Y)\ge 0$ and 
since $X$ is independent of $K$, then
$I(X;Y,V|K)=I(X;Y,V,K)\ge I(X;Y)$. Thus, for $Z=Y$, the achievable region
of Theorem 3 is a subset of the one given in Theorem 2. However, since
all these inequalities become equalities
at the same time by choosing $V=Y$ and letting both be independent of $K$,
the two regions are identical in the attack--free case.

Returning now to Theorem 3, as we observed,
$K^n$ is now involved not only in the role 
of a cipher key, but also as SI available at both encoder and decoder.
Two important points are now in order, in view of this fact.

First, one may argue that, actually,
there is no real reason to assume that $K^n$ is necessarily independent
of $X^n$ (see also \cite{MO03}).
If the user has control of the mechanism of generating the key,
then s/he might implement, in general, a channel 
$P_{K^n|X^n}(k^n|x^n)$ using the
available randomness resources, and taking (partial) advantage of the
randomness of the covertext. Let us 
assume that this channel is stationary and memoryless, i.e.,
\begin{equation}
P_{K^n|X^n}(k^n|x^n)=\prod_{t=1}^n P_{K|X}(k_t|x_t)
\end{equation}
with the single--letter transition probabilities
$\{P_{K|X}(k|x)~x\in\calX,~k\in\calK\}$ left as a degree of freedom
for design. While so far, we assumed that $K$ was independent of $X$,
the other extreme is, of course, $K=X$ (corresponding to private
watermarking). Note, however, that in the attack--free case, in the absence of
the compressibility requirement no.\ 2 (say, $R_c=\infty$), no optimality
is lost by assuming that $K$ is independent of $X$, since the only
inequality where we have used the independence assumption, in the previous
paragraph, corresponds to condition (d).

The second point is that in Theorems 1--3, so far, we have defined
the compressibility of the stegotext in terms of $H(Y^n)$, which is
suitable when the decompression of $Y^n$ is {\it public}, i.e., without
access to $K^n$. The legitimate decoder in our model, on the other hand,
has access to the SI $K^n$, which may depend on $Y^n$. In this context,
it then makes sense
to measure the compressibility of the stegotext 
also in a {\it private} regime,
i.e., in terms of the {\it conditional} entropy, $H(Y^n|K^n)$.

Our last (and most general) version of the 
coding theorem below takes these two points in to account.
Specifically, let us impose, in requirement no.\ 2, an additional inequality,
\begin{equation}
H(Y^n|K^n)\le nR_c', 
\end{equation}
where $R_c'$ is a prescribed constant, and let us redefine accordingly
the block codes and the achievable region in terms of six--tuples 
$(D,D',R_c,R_c',h,h')$. We now have the following result:

\begin{theorem}
A six--tuple $(D,D',R_c,R_c',h,h')$ is achievable 
if and only if there exist RV's $V$ and $Y$ such that
$P_{KXVYZ}(k,x,v,y,z)=P_{XK}(x,k)P_{VY|KX}(v,y|k,x)P_{Z|Y}(z|y)$,
where the alphabet size of $V$ is bounded by $|\calV|\le
|\calK|\cdot|\calX|\cdot|\calY|+1$, and such that the following
conditions are all satisfied:
\begin{itemize}
\item [(a)] $h\le H(K|Y)/\lambda+H(U)-R_U(D')$.
\item [(b)] $h'\le H(K|Y)/\lambda$.
\item [(c)] $\lambda R_U(D')\le I(V;Z|K)-I(V;X|K)$.
\item [(d)] $R_c\ge \lambda R_U(D')+I(X;Y,V|K)+I(K;Y)$.
\item [(e)] $R_c'\ge \lambda R_U(D')+I(X;Y,V|K)$.
\item [(f)] $D \ge Ed(X,Y)$.
\end{itemize}
\end{theorem}
Note that
the additional condition, (e), is similar to condition (d) except for
the term $I(K;Y)$. Also, in the joint PMF of $(K,X,V,Y,Z)$
we are no longer assuming that $K$ and $X$ are independent.
It should be pointed out that
in the presence of the new requirement regarding $H(Y^n|K^n)$,
it is more clear now that introducing dependence of $(V,Y)$ upon $K$
is reasonable, in general.
In the case $K=X$,
that was mentioned earlier,
the term $I(V;X|K)$, in condition (c),
and the term $I(X;Y,V|K)$, in conditions (d) and (e), both vanish.
Thus, both embedding performance and compression 
performance improve, like in private watermarking.

Finally, a comment is in order regarding the assumption $H(K)\le\lambda R_U(D')$,
which implies that $H(K|Y)$ cannot exceed $\lambda R_U(D')$ either.
If this assumption is removed, and even $H(K|Y)$ is allowed to exceed $\lambda R_U(D')$,
then Theorem 4 can be somewhat further extended. While $h$ cannot be further improved
if $H(K|Y)$ is allowed to exceed $\lambda R_U(D')$ (as it already reaches the maximum possible
value, $h=H(U)$, for $H(K|Y)=\lambda R_U(D')$), it turns out that there is still room for
improvement in $h'$. Suppose that instead of one rate--distortion codebook for $U^N$, we have
many {\it disjoint} codebooks. In fact, it has been shown in \cite{MM03} that there are exponentially
$2^{NH(\hU|U)}$ disjoint codebooks, 
each covering the set of typical source sequences by jointly typical
codewords. Now, if $H(K|Y) > \lambda R_U(D')$, we can use the $T=nH(K|Y)-NR_U(D')$ excess bits
of the compressed key
(beyond the $NR_U(D')$ bits that are used to encrypt the 
binary of representation of $\hU^N$), so as to select one of
$2^T$ codebooks (as long as $T < NH(\hU|U)$), and thus reach a total equivocation of $nH(K|Y)$ as long
as $nH(K|Y)\le NH(\hU)$, or equivalently, $H(K|Y)\le\lambda H(\hU)$. The equivocation level 
$h'=H(\hU)$ is now the ``saturation value'' that cannot be further improved (in analogy to $h=H(U)$
for the original source). This means that
condition (b) of Theorem 4 would now be replaced by the condition
\begin{equation}
\label{13}
h'\le \min\{H(\hU),H(K|Y)/\lambda\}.
\end{equation}
But with this condition, it is no longer clear that the best test channel for lossy compression of
$U^N$ is the one that achieves $R_U(D')$, because for the above modified version of condition (b),
it would be best to have $H(\hU)$ as large as possible (as long as it is below $H(K|Y)/\lambda$),
which is in partial conflict with the minimization 
of $I(U;\hU)$ that leads to $R_U(D')$. Therefore, a restatement of Theorem 4 would
require the existence of a channel $\{P_{\hU|U}(\hu|u),~u\in\calU,~\hu\in\hat{\calU}\}$ (in addition to
the existing requirement of a channel $P_{VY|KX}$), such that
the random variable $\hU$ takes now part in the 
compromise among {\it all} criteria of the problem. This means
that in conditions (a),(c),(d), and (e) of Theorem 4, 
$R_U(D')$ should be replaced by $I(U;\hU)$, and there would be an
additional condition (g): $Ed'(U,\hU)\le D'$. Condition (a), in view of the earlier discussion above,
would now be of the form:
\begin{equation}
\label{14}
h\le \min\{H(U),H(K|Y)/\lambda+H(U)-I(U;\hU)\}\equiv H(U)-[I(U;\hU)-H(K|Y)/\lambda]_+,
\end{equation}
where $[z]_+\dfn\max\{0,z\}$. 
Of course, under the assumption $H(K)\le\lambda R_U(D')$, that we have used thus far, 
\begin{equation}
H(\hU)\ge I(U;\hU)\ge R_U(D')\ge H(K)/\lambda\ge H(K|Y)/\lambda,
\end{equation}
in other words, $\min\{H(\hU),H(K|Y)/\lambda\}$ is always
attained by $H(K|Y)/\lambda$, and so, the dependence on $H(\hU)$ disappears, which means that the best
choice of $\hU$ (for all other conditions) is back to be the one that minimizes $I(U;\hU)$,
which gives us Theorem 4 as is. 

It is interesting to point out that 
this additional extension gives rise to yet 
another step in the direction of invalidating
the separation principle: While in Theorem 4 only the encryption and the
embedding interacted, yet the rate--distortion coding of $U^N$ was still
independent of all other ingredients
of the system, here even this is no longer true, as the choice
of the test channel $P_{\hU|U}$ takes into account also compromises that
are associated with the encryption and the embedding. 

Note that this discussion
applies also to the {\it classical} joint source--channel coding, where there is no
embedding at all: In this case, $X$ is a degenerate RV (say, $X\equiv 0$, if $0\in\calX$), and so,
the mutual information terms depending on $X$ in conditions (c), (d) and (e), all
vanish, the best choice of $V$ is $V=Y$ (thus, the r.h.s in condition (c) becomes the capacity
of the channel $P_{Z|Y}$ with $K$ as SI at both ends), 
and condition (f) may be interpreted as a (generalized) power
constraint (with power function $\phi(y)=d(0,y)$). Nonetheless, the new versions of conditions (a) and (b)
remain the same as in eqs.\ (\ref{13}) 
and (\ref{14}). This is to say that the violation of the separation principle
occurs even in the classical model of a communication system, 
once security becomes an issue
and one is interested in the security of the reconstructed source. 

\section{Proof of the Converse Part of Theorem 4}

Let an $(n,\lambda,D+\epsilon,D'+\epsilon,R_c+\epsilon,
R_c'+\epsilon,h-\epsilon,h'-\epsilon)$ code be given.
First, from the requirement $H(Y^n|K^n)\le n (R_c'+\epsilon)$, we have:
\begin{eqnarray}
n(R_c'+\epsilon) &\ge& H(Y^n|K^n)\label{1st}\\
&=&H(Y^n|U^N,K^n)+I(U^N;Y^n|K^n)\nonumber\\
&\ge&H(Y^n|U^N,K^n)+I(U^N;Z^n|K^n)\nonumber\\
&=&H(Y^n|U^N,K^n)+I(U^N;Z^n,K^n) \label{2nd}
\end{eqnarray}
where the second inequality comes from the data processing theorem
($U^N\to Y^n\to Z^n$ is a Markov chain given $K^n$)
and the last equality comes from the 
chain rule and the fact that $U^N$ and $K^n$
are independent. Define
$\tilde{V}_t=(X_{t+1}^n,U^N,K^{t-1},Z^{t-1})$, 
$J$ -- as a uniform RV over $\{1,\ldots,n\}$, $X=X_J$, $K=K_J$, $Y=Y_J$,
$V'=\tilde{V}_J$, and $V=(\tilde{V}_J,J)=(V',J)$.
Now, the first term on the 
right--most side of eq.\ (\ref{2nd}) is further lower bounded 
in the following manner. 
\begin{eqnarray}
H(Y^n|U^N,K^n)&\ge&I(X^n;Y^n|U^N,K^n)\nonumber\\
&=&I(X^n;Y^n,U^N,K^n)-I(X^n;U^N,K^n)\nonumber\\
&=&\sum_{t=1}^n I(X_t;Y^n,U^N,K^n|X_{t+1}^n)-I(X^n;K^n)\label{xp1}\\
&=&\sum_{t=1}^n I(X_t;Y^n,U^N,K^n,X_{t+1}^n)-nI(X;K)\label{xp2}\\
&\ge&\sum_{t=1}^n 
I(X_t;K_t,Y_t,U^N,K^{t-1},Z^{t-1},X_{t+1}^n)-nI(X;K)\label{xp3}\\
&=&\sum_{t=1}^n I(X_t;K_t,Y_t,\tilde{V}_t)-nI(X;K)\nonumber\\
&=&n[I(X;K,Y,V'|J)-I(X;K)]\nonumber\\
&=&n[I(X;K,Y,V',J)-I(X;K)]\label{xp4}\\
&=&nI(X;Y,V|K)\label{4th}
\end{eqnarray}
where (\ref{xp1}) is due to the chain 
rule and fact that $(X^n,K^n)$ is independent of
$U^N$ (hence $U^N\to K^n\to X^n$ is trivially a Markov chain),
(\ref{xp2}) is due to the memorylessness of $\{(X_t,K_t)\}$,
(\ref{xp3}) is due to the data processing theorem,
and (\ref{xp4}) follows from the fact that $\{X_t\}$ is stationary
and so, $X=X_J$ is independent of $J$.
The second term on the right--most 
side of eq.\ (\ref{2nd}) is in turn lower bounded following
essentially the same ideas as in the proof of the converse
to the rate--distortion coding theorem (see, e.g., \cite{CT91}):
\begin{eqnarray}
I(U^N;Z^n,K^n)&=&H(U^N)-H(U^N|Z^n,K^n)\nonumber\\
&=&\sum_{i=1}^N[H(U_i)-H(U_i|U^{i-1},Z^n,K^n)]\nonumber\\
&=&\sum_{i=1}^N I(U_i;U^{i-1},Z^n,K^n)\nonumber\\
&\ge&\sum_{i=1}^N I(U_i;[g_n(Z^n,K^n)]_i)\nonumber\\
&\ge&\sum_{i=1}^N R_U(Ed'(U_i,[g_n(Z^n,K^n)]_i))\nonumber\\
&\ge&NR_U\left(\frac{1}{N}\sum_{i=1}^N
Ed'(U_i,[g_n(Z^n,K^n)]_i)\right)\nonumber\\
&\ge&NR_U(D'+\epsilon),\label{5th}
\end{eqnarray}
where $[g_n(Z^n,K^n)]_i$ denotes the $i$-th component projection
of $g_n(Z^n,K^n)$, i.e., $\hU_i$ as a function of $(Z^n,K^n)$.
Combining eqs.\ (\ref{2nd}), (\ref{4th}), and (\ref{5th}), we get
\begin{equation}
n(R_c'+\epsilon)\ge NR_U(D'+\epsilon)+nI(X;Y,V|K).
\end{equation}
Dividing by $n$, we get 
\begin{equation}
\label{6th}
R_c'+\epsilon\ge \lambda R_U(D'+\epsilon)+I(X;Y,V|K).
\end{equation}
Using the arbitrariness of $\epsilon$ together
with the continuity of 
$R_U(\cdot)$, we get condition (e) of Theorem 4.

Condition (d) is derived in the very same manner except that
the starting point is the inequality $n(R_c+\epsilon)\ge H(Y^n)$,
and when $H(Y^n)$ is further bounded from below, in analogy
to the chain of inequalities (\ref{2nd}), there is 
an additional term, $I(K^n;Y^n)$,
that is in turn
lower bounded in the following manner:
\begin{eqnarray}
I(K^n;Y^n)&\ge&\sum_{t=1}^n I(K_t;Y_t)\nonumber\\
&=&nI(K;Y|J)\nonumber\\
&=&n[H(K|J)-H(K|J,Y)]\nonumber\\
&\ge&n[H(K)-H(K|Y)]\nonumber\\
&=&nI(K;Y),
\end{eqnarray}
where the first inequality is because of the memorylessness of $\{K_t\}$,
and the second inequality comes from the facts that 
conditioning reduces entropy (in the second term) 
and that $K$ is independent of
$J$ (again, due to the stationarity of $\{K_t\}$).
This gives the additional term, $I(K;Y)$, in condition (d).

Condition (c) is obtained as follows:
\begin{eqnarray}
NR_U(D'+\epsilon)&\le&I(U^N;K^n,Z^n)\nonumber\\
&=&I(U^N;K^n,Z^n)-I(U^N;K^n,X^n)\nonumber\\
&\le&\sum_{t=1}^n[I(\tilde{V}_t;K_t,Z_t)-I(\tilde{V}_t;K_t,X_t)]\\
&=&n[I(V';K,Z|J)-I(V';K,X|J)]\nonumber\\
&\le&n[I(V',J;K,Z)-I(V',J;K,X)]\\
&=&n[I(V;K,Z)-I(V;K,X)]\nonumber\\
&=&n[I(V;Z|K)-I(V;X|K)],
\end{eqnarray}
where 
the first inequality is (\ref{5th}),
the first equality is due to the independence between $U^N$ and $(K^n,X^n)$,
the second inequality is 
an application of \cite[Lemma 4]{GP80},
the third inequality is due to the fact 
that $I(K,Z;J)\ge 0$ and $I(K,X;J)=0$ (due to the
stationarity of $\{(K_t,X_t)\}$), and 
the last equality is obtained by adding and subtracting
$I(V;K)$. Again, since this is true for every $\epsilon > 0$,
it holds also for $\epsilon=0$, due to continuity. 

As for condition (f), we have:
\begin{equation}
D+\epsilon\ge\frac{1}{n}\sum_{t=1}^nEd(X_t,Y_t)=Ed(X,Y),
\end{equation}
and we use once again the arbitrariness of $\epsilon$.
Regarding condition (b), we have:
\begin{eqnarray}
nH(K|Y)&\ge&nH(K|Y,J)\nonumber\\
&=&\sum_{t=1}^nH(K_t|Y_t)\nonumber\\
&\ge&\sum_{t=1}^nH(K_t|K^{t-1},Y^n)\nonumber\\
&=& H(K^n|Y^n)\nonumber\\
&=& H(K^n|Y^n,Z^n)\nonumber\\
&\ge&I(K^n;\hat{U}^N|Y^n,Z^n)\nonumber\\
&=&H(\hat{U}^N|Y^n,Z^n)-H(\hat{U}^N|Y^n,Z^n,K^n)\nonumber\\
&=&H(\hat{U}^N|Y^n,Z^n)\nonumber\\
&\ge&N(h'-\epsilon),
\end{eqnarray}
where the last equality is due to the fact that $\hat{U}^N$
is, by definition, a function of $(Z^n,K^n)$, and the last
inequality is by the hypothesis that the code achieves an equivocation of
at least $N(h'-\epsilon)$. Dividing by $N$ 
and taking the limit $\epsilon\to 0$,
leads to  
$h'\le H(K|Y)/\lambda$, which is condition (b).
Finally, to prove condition (a), consider the inequality 
$nH(K|Y)\ge H(\hat{U}^N|Y^n,Z^n)$, that we have just proved,
and proceed as follows (see also \cite{Yamamoto97}):
\begin{eqnarray}
\label{8th}
nH(K|Y)&\ge&H(\hat{U}^N|Y^n,Z^n)\nonumber\\
&\ge&H(\hat{U}^N|Y^n,Z^n)+N(h-\epsilon)-H(U^N|Y^n,Z^n)\nonumber\\
&=&N(h-\epsilon)-H(U^N)+I(U^N;Y^n,Z^n)-\nonumber\\
& &I(\hat{U}^N;Y^n,Z^n)+I(\hat{U}^N;U^N)+H(\hat{U}^N|U^N)\nonumber\\
&\ge&N[h-\epsilon-H(U)+R_U(D'+\epsilon)]+\nonumber\\
& &[I(U^N;Y^n,Z^n)-I(\hat{U}^N;Y^n,Z^n)+H(\hat{U}^N|U^N)],
\end{eqnarray}
where the second inequality follows from the hypothesis that
the code satisfies $H(U^N|Y^n,Z^n)\ge N(h-\epsilon)$,
and the third inequality is due to the memorylessness of $\{U_i\}$, the
hypothesis that $\sum_{i=1}^NEd'(U_i,\hU_i)\le N(D'+\epsilon)$,
and the converse to the rate--distortion coding theorem.
Now, to see that the second bracketed term is non--negative, we have the
following chain of inequalities:
\begin{eqnarray}
\label{bra}
&& I(U^N;Y^n,Z^n)-I(\hat{U}^N;Y^n,Z^n)+H(\hat{U}^N|U^N)\nonumber\\
&=&I(U^N;Y^n,Z^n)-H(Y^n,Z^n)+H(Y^n,Z^n|\hat{U}^N)
+H(\hat{U}^N|U^N)\nonumber\\
&\ge& I(U^N;Y^n,Z^n)-H(Y^n,Z^n)+H(Y^n,Z^n|U^N,\hat{U}^N)
+H(\hat{U}^N|U^N)\nonumber\\
&=& I(U^N;Y^n,Z^n)-H(Y^n,Z^n)+H(Y^n,Z^n,\hat{U}^N|U^N)\nonumber\\
&\ge& I(U^N;Y^n,Z^n)-H(Y^n,Z^n)+H(Y^n,Z^n|U^N)\nonumber\\
&=& 0.
\end{eqnarray}
Combining this with eq.\ (\ref{8th}), we have
\begin{equation}
nH(K|Y)\ge
N[h-\epsilon-H(U)+R_U(D'+\epsilon)].
\end{equation}
Dividing again by $N$, and letting $\epsilon$ vanish, we obtain 
$h\le H(K|Y)/\lambda+H(U)-R_U(D')$, which 
completes the proof of condition (a).

To complete the proof of the converse part, it remains to show that the
alphabet size of $V$ can be reduced to $|\calK|\cdot|\calX|\cdot|\calY|+1$.
To this end, we extend the proof of the parallel argument in \cite{MM04}
by using the support lemma (cf.\ \cite{CK81}), which is based on
Carath\'{e}odory's theorem. According to this lemma, given $J$ real
valued continuous functionals $f_{j}$, $j=1,...,J$ on the set
$\calP(\calX)$ of probability distributions over the alphabets
$\calX$, and given any probability measure $\mu$ on the Borel
$\sigma$-algebra of $\calP(\calX)$, there exist $J$ elements
$Q_{1},...,Q_{J}$ of $\calP(\calX)$ and $J$ non-negative reals,
$\alpha_{1},...,\alpha_{J}$, such that
$\sum_{j=1}^{J}\alpha_{j}=1$ and for every $j=1,...,J$
\begin {eqnarray}
    \int_{\calP(\calX)}f_{j}(Q)\mu(dQ) =
    \sum_{i=1}^{J}\alpha_{i}f_{j}(Q_{i}).
\end {eqnarray}
Before we actually apply the support lemma, we first rewrite the
relevant mutual informations of Theorem 4 in a more convenient
form for the use of this lemma. First, observe that
\begin {eqnarray}
    I(V;Z|K)-I(V;X|K)   & = & H(Z|K)-H(Z|V,K) - H(X|K) + H(X|V,K)\nonumber\\
                        & = & H(Z|K)-H(X|K) + H(K,X|V)-H(K,Z|V).
\end {eqnarray}
and 
\begin {eqnarray}
I(X;Y,V|K)  & = & I(X;V|K) + I(X;Y|V,K) \\
                        & = & H(X|K) - H(X|V,K) +
                        H(X|V,K)-H(X|V,Y,K) \nonumber \\
                        & = & H(X|K)-H(X|V,Y,K) \nonumber \\
                        & = & H(X|K)-H(K,X,Y|V)+H(K,Y|V).
\end {eqnarray}
For a given joint distribution of 
$(K,X,Y)$, and given $P_{Z|Y}$, $H(Z|K)$ and $H(X|K)$ are
both given and unaffected by $V$. Therefore, in order to preserve
prescribed values of $I(V;Z|K)-I(V;X|K)$ and $I(X;V,Y|K)$, it is
sufficient to preserve the associated values $H(K,X|V) - H(K,Z|V)$
and $H(K,X,Y|V) - H(K,Y|V)$.
Let us define then the following functionals of a generic
distribution $Q$ over $\calK\times\calX\times\calY$, 
where $\calK\times\calX \times
\calY$ is assumed, without loss of generality, to be
$\{1,2,...,m\}$, $m = |\calK|\cdot|\calX|\cdot|\calY|$:
\begin{eqnarray}
    &&f_{i}(Q) = Q(k,x,y), ~~~i\dfn (k,x,y)=1,...,m-1\\
    &&f_{m}(Q) =
    \sum_{k,x,y}Q(k,x,y)\sum_{z}P_{Z|Y}(z|y)
\log\frac{\sum_{x,y}Q(k,x,y)P_{Z|Y}(z|y)}{Q(k,x)}.
\end{eqnarray}
Next define
\begin{eqnarray}
    &&f_{m+1}(Q) =
    \sum_{k,x,y}Q(k,x,y)\log\frac{Q(k,y)}{Q(k,x,y)}.
\end{eqnarray}
Applying now the support lemma, we find that there exists a random
variable $V$ (jointly distributed with $(K,X,Y)$), whose alphabet
size is $|\calV| = m+1 = |\calK|\cdot|\calX|\cdot|\calY|+1$ and it satisfies
simultaneously:
\begin{eqnarray}
    \sum_{v}\Pr\{V = v\}f_{i}(P(\cdot|v)) = P_{KXY}(k,x,y), \textrm{ }
    i=1,...,m-1,
\end{eqnarray}
\begin{eqnarray}
    \sum_{v}\Pr\{V = v\}f_{m}(P(\cdot|v)) = H(K,X|V) - H(K,Z|V),
\end{eqnarray}
and
\begin{eqnarray}
    \sum_{u}\Pr\{V = v\}f_{m+1}(P(\cdot|v)) = H(K,X,Y|V) -
    H(K,Y|V).
\end{eqnarray}
It should be pointed out that this random variable maintains the
prescribed distortion level $Ed(X,Y)$ since 
$P_{XY}$ is preserved. By the same token, $H(K|Y)$ and $I(K;Y)$, which depend
only on $P_{KY}$, are preserved as well.
This completes the proof of the
converse part of Theorem 4.

\section{Proof of the Direct Part of Theorem 4}

In this section, we show that if there exist RV's $(V,Y)$ that satisfy the
conditions of Theorem 4, then for every $\epsilon > 0$,
there is a sufficiently large $n$ for which 
$(n,\lambda,D+\epsilon,D'+\epsilon,R_c+\epsilon,
R_c'+\epsilon,h-\epsilon,h'-\epsilon)$ codes exist. 
One part of the proof is strongly based
on a straightforward extension of the proof of the direct part of 
\cite{MM04} to the case 
of additional SI present at both encoder and decoder. Nonetheless,
for the sake of completeness, the full details are provided here.
It should be pointed out that for the attack--free case, an analogous
extension can easily be offered to the direct part of \cite{MM03}.

We first digress to establish some additional notation
conventions associated with the method of types \cite{CK81}. For a
given generic 
finite--alphabet random variable (RV) $A \in \calA$ (or a vector of
RV's taking on values in $\calA$), and a vector $a^\ell \in
\calA^{\ell}$ ($\ell$ -- positive integer), 
the empirical probability mass function (EPMF) is a
vector $P_{a^\ell}=\{P_{a^\ell}(a'),~a' \in \calA\}$, where $P_{a^\ell}(a')$ is
the relative frequency of the letter $a' \in \calA$ in the vector
$a^\ell$. Given $\delta > 0$, let us denote the set of all
$\delta$-typical sequences of length $\ell$ by
$T_{P_A}^\delta$, or by $T_A^\delta$ 
(if there is no ambiguity regarding the 
PMF that governs $A$), i.e., $T_A^\delta$ 
is the set of the sequences $a^\ell \in
\calA^\ell$ such that
\begin{equation}
\label{Px}
    (1-\delta)P_{A}(a') \leq P_{a^\ell}(a') \leq (1+\delta)P_{A}(a')
\end{equation}
for every $a' \in \calA$. For sufficiently large $\ell$,
the size of $T_A^\delta$ is well--known \cite{CK81} to be bounded by
\begin {equation}
\label{TgxSize}
    2^{\ell[(1-\delta)H(A)-\delta]} \leq
    |T_A^\delta| \leq 2^{\ell(1+\delta)H(A)}.
\end{equation}
It is also well--known (by the weak law of large numbers)
that:
\begin{equation}
\label{PrTgx}
    \Pr \big\{ A^\ell \notin T_A^\delta \big\} \leq \delta
\end{equation}
for all $\ell$ sufficiently large.
For a given generic channel $P_{B|A}(b|a)$ and for each $a^\ell \in
T_A^\delta$, the set 
of all sequences $b^l$ that are jointly
$\delta$-typical with $a^\ell$, will be denoted by
$T_{P_{B|A}}^\delta(a^\ell)$, 
or by $T_{B|A}^\delta(a^\ell)$ if there is no ambiguity, i.e.,
$T_{B|A}^\delta(a^\ell)$ is the set of all $b^\ell$ such that:
\begin{equation}
\label{Pygx}
    (1-\delta)P_{a^\ell}(a')P_{B|A}(b'|a') \leq P_{a^\ell b^\ell}(a',b') \leq
    (1+\delta)P_{a^\ell}(a')P_{B|A}(b'|a'),
\end{equation}
for all $a'\in \calA, b'\in \calB$, where $P_{a^\ell b^\ell}(a',b')$
denotes the fraction of occurrences of the pair $(a',b')$ in
$(a^\ell,b^\ell)$. Similarly as in eq.\ (\ref{Px}), for all sufficiently large $\ell$ and
$a^\ell \in T_A^\delta$, the size of $T_{B|A}^\delta(a^\ell)$ is bounded as follows:
\begin {equation}
\label{TgyxSize}
    2^{\ell[(1-\delta)H(B|A)-\delta]} \leq
    |T_{B|A}^\delta(a^\ell)| \leq 2^{\ell(1+\delta)H(B|A)}.
\end{equation}
Finally, observe that for all $a^\ell \in T_A^\delta$ and $b^\ell
\in T_{B|A}^\delta(a^\ell)$, the distortion 
$d(a^\ell,b^\ell)=\sum_{j=1}^\ell d(a_j,b_j)$ is upper bounded by:
\begin{equation}
\label{d_theoretic}
    d(a^\ell,b^\ell) \leq \ell(1+\delta)^{2}\sum_{a',b'}P_{A}(a')P_{B|A}(b'|a')d(a',b') 
\dfn \ell(1+\delta)^{2}Ed(A,B).
\end{equation}

Let $(K,X,V,Y,Z)$ be a given random vector that 
satisfies the conditions of Theorem 4.
We now describe the mechanisms of random code selection
and the encoding and decoding operations. For a given $\epsilon > 0$, fix $\delta$ such that
$2\delta+\max\{2\cdot\exp\{-2^{n\delta}\}+2^{-n\delta},\delta^{2}\}
\leq \epsilon$. Define also
\begin{equation}
\epsilon_1\dfn \delta[1+H(V|K)+H(V|K,X)],
\end{equation}
\begin{equation}
\epsilon_2\dfn \delta[1+H(Y|K,V)+H(Y|K,X,V)],
\end{equation}
and 
\begin{equation}
\epsilon_3\dfn\delta[1+H(V|K)+H(V|Z,K)].
\end{equation}
\\
\\
\noindent \textsl{Generation of a rate--distortion code}: \\
Apply the type--covering lemma \cite{CK81} and
construct a rate--distortion codebook that covers $T_U^\delta$
within distortion 
$N(D'+\epsilon)$ w.r.t.\ $d'$, using $2^{NR_U(D')}$ codewords.
\\
\\
\noindent \textsl{Generation of the encrypting bitstream}: \\
For every $k^n\in T_K^\delta$, randomly select an index in the
set $\{0,1,\ldots,2^{n[H(K|Y)+\delta]}-1\}$ with a uniform
distribution. Denote by
$s^J(k^n)=(s_1(k^n),\ldots,s_J(k^n))$, $s_j(k^n)\in\{0,1\}$, $j=1,\ldots,J$, 
the binary string of length $J=n[H(K|Y)+\delta]$ that represents
this index. (Note that $s^J(k^n)$ can be interpreted as the output
of the Slepian--Wolf encoder for $K^n$, where $Y^n$ plays the role
of SI at the decoder \cite{SW73}.)
\\
\\
\noindent \textsl{Generation of an auxiliary embedding code}: \\
We first construct an auxiliary code capable of
embedding $2^{NR_U(D')}$ watermarks by a random selection technique.
First, $M_1=2^{nR_1}$, $R_1 = I(V;Z|K)-\epsilon_3-\delta$,
sequences $\{V^n(i,k^n)\}$, $i\in\{1,\ldots,M_1\}$, are drawn
independently from $T_{V|K}^\delta(k^n)$ for every $k^n\in T_{K}^\delta$. 
For every such $k^n$, let us denote the set of these
sequences by $\calC(k^n)$. The elements of $\calC(k^n)$ are evenly
distributed among $M_U \dfn 2^{NR_U(D')}$ bins, each of size $M_2
= 2^{nR_2}$, $R_2 = I(X;V|K)+\epsilon_1 + \delta$ (this is possible
thanks to condition (c) of Theorem 4, provided 
that the inequality therein is strict). A different
(encrypted) message of length $L=NR_U(D')=n\lambda R_U(D')$ bits
is attached to each bin, identifying a sub-code
that represents this message. We denote the codewords in bin number $m$ ($m
\in \{1,2,\ldots,M_U\}$), by $\{V^n(m,j,k^n)\}$, $j \in \{1,2,\ldots,M_2\}$.
\\
\\
\noindent \textsl{Stegotext sequence generation}: \\
\noindent For each auxiliary sequence (in the above auxiliary codebook
of each $\delta$--typical $k^n$),
$V^n(m,j,k^n)=v^n$, a set of $M_3 \dfn 2^{nR_3}$, $R_3 =
I(X;Y|V,K)+\epsilon_2 + \delta$, stegotext sequences
$\{Y^n(j',v^n,k^n)\}$, $j' \in \{1,\ldots,M_3\}$, are independently drawn from
$T_{Y|VK}^\delta(v^n,k^n)$. We denote this set by $\calC(v^n,k^n)$.
\\
\\
\noindent \textsl{Encoding}:\\
\noindent Upon receiving a triple $(u^N,x^n,k^n)$, the encoder acts as
follows:
\begin{enumerate}
\item If $u^N\in T_U^\delta$, 
let $w^L=(w_1,\ldots,w_L)$, $w_i\in\{0,1\}$, $i=1,\ldots,L$
be the binary representation of the index of the rate--distortion
codeword for the message source. 
For $k^n\in T_K^\delta$, let $s^J(k^n)=(s_1(k^n),\ldots,s_J(k^n))$
denote binary representation string of the index of $k^n$.
Let $\tilde{w}^L=(\tilde{w}_1,\ldots,\tilde{w}_L)$, where
$\tilde{w}_j=w_j\oplus s_j(k^n)$, 
$j=1,\ldots,J$, and $\tilde{w}_j=w_j$, $j=J+1,\ldots,L$,
and where $\oplus$ denotes modulo 2
addition i.e., the XOR operation.\footnote{Note that since $H(K)$ is 
assumed smaller than $\lambda R_U(D')$,
then so is $H(K|Y)$,
and therefore $J\le L$.} 
The binary vector $\tilde{w}^L$ is the (partially) encrypted message to be
embedded. Let $m=\sum_{l=1}^L\tilde{w}_l2^{l-1}+1$ 
denote the index of this message.
If $u^N\notin T_{U}^\delta$ or $k^n\notin T_{K}^\delta$, an arbitrary
(error) message $\tilde{w}^L$ is generated (say, the all--zero message).
\item If $(k^n,x^n) \in T_{KX}^\delta$ 
find, in bin number $m$, the first $j$ such that
$V^n(m,j,k^n)=v^n$ is jointly typical, i.e.,
$(k^n,x^n,v^n) \in T_{KXV}^\delta$, and then find the 
first $j'$ such that $Y^n(j',v^n,k^n)=y^n 
\in \calC(v^n,k^n)$ is jointly typical, i.e.,
$(k^n,x^n,v^n,y^n) \in T_{KXVY}^\delta$. 
This vector $y^n$ is chosen for transmission.
If $(k^n,x^n) \notin T_{KX}^\delta$, 
or if there is no $V^n(m,j,k^n)=v^n$ and $Y^n(j',v^n,k^n)=y^n$ such that
$(k^n,x^n,v^n,y^n) \in T_{KXVY}^\delta$, 
an arbitrary vector $y^n\in\calY^n$ is transmitted.
\end {enumerate}

\noindent \textsl{Decoding}:\\
\noindent Upon receiving $Z^n = z^n$ 
and $K^n=k^n$, the decoder finds all
sequences $\{v^n\}$ in $\calC(k^n)$ such that $(k^n,v^n,z^n) \in
T_{KVZ}^\delta$. If all $\{v^n\}$ 
found belong to the same bin, say, $\hat{m}$,
then $\hat{m}$ is decoded as the 
embedded message, and then the binary representation
vector $\hat{w}^L=(\hat{w}_1,\ldots,\hat{w}_L)$ corresponding to $\hat{m}$ is 
decrypted, again, by modulo 2 addition
of its first $J$ bits with $s^J(k^n)$. 
This decrypted binary $L$--vector is
then mapped to the corresponding reproduction 
vector $\tilde{u}^N$ of the rate--distortion codebook
for the message source.
If there is no $v^n\in\calC(k^n)$ 
such that $(k^n,v^n,z^n) \in T_{KVZ}^\delta$ or if
there exist two or more bins that contain such a sequence, an
error is declared.
\\
\\
\noindent We now turn to the performance 
analysis of this code in all relevant aspects.
For each triple $(k^n,x^n,u^N)$ and particular choices of
the codes, the possible causes for incorrect
watermark decoding are the following:

\begin{enumerate}
  \item $(k^n,x^n,u^N) \notin T_{KX}^\delta\times T_{U}^\delta$.
   Let the probability of this event be defined as $P_{e_{1}}$.
  \item $(k^n,x^n,u^N) 
    \in T_{KX}^\delta\times T_{U}^\delta$, but in bin no.\ $m$ 
    there is no $v^n$ s.t. $(k^n,x^n,v^n) \in T_{KXV}^\delta$.
    Let the probability of this event be defined as $P_{e_{2}}$.
  \item $(k^n,x^n,u^N) 
    \in T_{KX}^\delta\times T_{U}^\delta$ and in bin no.\ $m$ 
    there is $v^n$ s.t. $(k^n,x^n,v^n) \in T_{KXV}^\delta$, but there is no
    $y^n \in \calC(v^n,k^n)$ s.t. $(k^n,x^n,v^n,y^n) \in T_{KXVY}^\delta$.
    Let the probability of this event be defined as $P_{e_{3}}$.
  \item $(k^n,x^n,u^N) 
    \in T_{KX}^\delta\times T_{U}^\delta$ and in bin no.\ $m$ 
    there is $v^n$ and $y^n
    \in \calC(v^n,k^n)$ such that $(k^n,x^n,v^n,y^n) \in T_{KXVY}^\delta$,
    but $(k^n,v^n,z^n) \notin T_{KVZ}^\delta$.
    Let the probability of this event be defined as $P_{e_{4}}$.
  \item $(k^n,x^n,u^N) 
    \in T_{KX}^\delta\times T_{U}^\delta$ and in bin no.\ $m$ 
    there is $v^n$ and $y^n
    \in \calC(v^n,k^n)$ such that $(k^n,x^n,v^n,y^n) \in T_{KXVY}^\delta$,
    and $(k^n,v^n,z^n) \in T_{KVZ}^\delta$, but
    there exists another bin, say, no.\ $\tilde{m}$,
    that contains $\tilde{v}^n$ s.t. $(k^n,\tilde{v}^n,z^n)\in T_{KVZ}^\delta$.
    Let the probability of this event be defined as $P_{e_{5}}$.
\end{enumerate}
If none of these events occur, the message 
$\tilde{w}^L$ (or, equivalently, $m$) is decoded
correctly from $z^n$, the distortion constraint between $x^n$
and $y^n$ is within $n(D+\epsilon)$ (as follows from (\ref{d_theoretic})), and
the distortion between $u^N$ and its rate--distortion codeword,
$\tilde{u}^N=\hat{u}^N$, does not exceed $N(D'+\epsilon)$. 
Thus, requirements 1 and 4 (modified according
to eq.\ (\ref{distp}), with $D'+\epsilon$ replacing $D'$) are both satisfied.
Therefore, we first prove that the probability for none of the
events 1--5 to occur, tends to unity as $n\to\infty$.

The average probability of error $P_{e}$ in decoding $m$ is bounded by
\begin {equation}
\label{Pe}
    P_{e} \leq \sum_{i=1}^5 P_{e_i}.
\end {equation}
The fact that $P_{e_{1}}\rightarrow 0$ follows immediately from
(\ref{PrTgx}). As for $P_{e_{2}}$, we have:
\begin {equation}
\label{P_e2a}
P_{e_{2}} \dfn \prod_{j=1}^{M_2}
\Pr\{(k^n,x^n,V^n(m,j,k^n)) \notin T_{KXV}^\delta\}.
\end {equation}
Now, by (\ref{TgxSize}), for every $j$ and every 
$(k^n,x^n)\in T_{KX}^\delta$:
\begin {eqnarray}
\label{P_e2i}
    \Pr\{V^n(m,j,k^n) \notin T_{V|KX}^\delta(k^n,x^n)\} & = & 1 -
\Pr\{V^n(m,j,k^n) \in T_{V|KX}^\delta(k^n,x^n)\} \nonumber\\
& = & 1 - \frac{|T_{V|KX}^\delta(k^n,x^n)|}{|T_{V|K}^\delta(k^n)|} \nonumber\\
    & \leq & 1 -
\frac{2^{n[(1-\delta)H(V|K,X)-\delta]}}
{2^{n(1+\delta)H(V|K)}} \nonumber\\
    & = &1 - 2^{-n[I(X;V|K)+\epsilon_1]}.
\end {eqnarray}
Substitution of
(\ref{P_e2i}) into (\ref{P_e2a}) provides us with the following
upper bound:
\begin {equation}
P_{e_{2}} \leq \Big[1 - 2^{-n[I(X;V|K)+\epsilon_1]}\Big]^{M_2} \leq
\exp\bigg\{-2^{nR_2}\cdot2^{-n[I(X;V|K)+\epsilon_1]}\bigg\} \rightarrow 0,
\end {equation}
double--exponentially rapidly since $R_2 = I(X;V|K)+\epsilon_1 +
\delta$.
To estimate $P_{e_{3}}$, we repeat the same technique:
\begin {equation}
\label{P_e3a}
P_{e_{3}} \dfn \prod_{j'=1}^{M_3}\Pr\{(k^n,x^n,v^n,Y^n(j',v^n,k^n))
\notin T_{KXVY}^\delta\}.
\end {equation}
Again, by the property of the typical sequences, for every $j'$ and 
$(k^n,x^n,v^n)\in T_{KXV}^\delta$:
\begin {eqnarray}
\label{P_e3Cu}
    \Pr\{Y^n(j',v^n,k^n) \notin T_{Y|KXV}^\delta(k^n,x^n,v^n)\} 
\leq 1 - 2^{-n[I(X;Y|V,K)+\epsilon_2]},
\end {eqnarray}
and therefore,
substitution of (\ref{P_e3Cu}) into (\ref{P_e3a}) gives
\begin {equation}
P_{e_{3}} \leq \Big[1 - 2^{-n[I(X;Y|V,K)+\epsilon_2]}\Big]^{M_3} \leq
\exp\bigg\{-2^{nR_3}\cdot2^{-n[I(X;Y|V,K)+
\epsilon_2]}\bigg\} \rightarrow 0,
\end {equation}
double--exponentially rapidly since $R_3 =
I(X;Y|V,K)+\epsilon_2 + \delta$.
The estimation of $P_{e_{4}}$ is again based on properties
of typical sequences. Since $Z^n$ is the output of
a memoryless channel $P_{Z|Y}$ with input
$y^n=Y^n(j',v^n,k^n)$ and by the assumption of this step
$(k^n,x^n,v^n,y^n) \in T_{KXVY}^\delta$, from (\ref{PrTgx}) 
and the Markov lemma \cite[Lemma 14.8.1]{CT91}, we obtain
\begin {equation}
\label{Pe4}
    P_{e_{4}} = \Pr\{(k^n,x^n,v^n,y^n,Z^n)
                \notin T_{KXVYZ}^\delta\} \leq \delta,
\end {equation}
and similarly to $P_{e_{1}}$, $P_{e_4}$ can be made as small as desired by an
appropriate choice of $\delta$.

Finally, we estimate $P_{e_{5}}$ as follows:
\begin {eqnarray}
    P_{e_{5}} & = & \Pr\{\exists \tilde{m} \neq m:
        (k^n,V^n(\tilde{m},j,k^n),z^n) \in T_{KVZ}^\delta\} \\
    & \leq & \sum_{\tilde{m} \neq m,~j\in\{1,2,...,M_2\}}
\Pr\{(k^n,V^n(\tilde{m},j,k^n),z^n) \in T_{KVZ}^\delta\}
        \nonumber \\
&=& (2^{NR_U(D')}-1)2^{nR_2}\Pr\{(k^n,V^n(\tilde{m},j,k^n),z^n) 
\in T_{KVZ}^\delta\} 
\nonumber \\
    & \leq & 2^{nR_1}2^{-n[I(V;Z|K)-\epsilon_3]}.
\end {eqnarray}
Now, since $R_1
= I(V;Z|K)-\epsilon_3-\delta$, $P_{e_{5}} \rightarrow 0$.
Since $P_{e_{i}}\rightarrow 0$ for $i=1,\ldots,5$,
their sum tends to zero as well, implying that there exist at
least one choice of an auxiliary code and related stegotext codes
that give rise to the reliable 
decoding of $\tilde{W}^L$.
 
Now, let us denote by $N_{c}$ the total number of composite
sequences in a codebook that corresponds 
to a $\delta$--typical $k^n$. Then,
\begin {eqnarray}
N_c&=&M_U\cdot M_2\cdot M_3\nonumber\\
&=&2^{n[\lambda R_U(D')+I(X;V|K)+I(X;Y|V,K)
+\epsilon_1+\epsilon_2+2\delta]}\nonumber\\
&=&2^{n[\lambda R_U(D')+I(X;Y,V|K)+\epsilon_1+\epsilon_2+2\delta]}.
\end {eqnarray}
Thus, 
\begin{eqnarray}
H(Y^n|K^n)&\leq&\log N_c\nonumber\\
&=&n[\lambda R_U(D')+I(X;Y,V|K)+\epsilon_1+\epsilon_2+2\delta]\nonumber\\
&\le& n(R_c'+\epsilon_1+\epsilon_2+2\delta),
\end{eqnarray}
where in the last inequality we have used condition (e).
For sufficiently small values of $\delta$ (and hence of $\epsilon_1$ and $\epsilon_2$)
$\epsilon_1+\epsilon_2+2\delta\le \epsilon$ and so, the compressibility
requirement in the presence of $K^n$ is satisfied.

We next prove the achievability of $R_c$. Let us consider the set of 
$\delta$--typical key sequences $T_K^\delta$, and view it as the
union of $0$--typical sets
(i.e., $\delta$--typical sets
with $\delta=0$),
$\{T_{Q_K}^0\}$, where $Q_K$ exhausts the set of all rational PMF's
with denominator $n$, and with the property
\begin{equation}
(1-\delta)P_K(k)\le Q_K(k)\le (1+\delta)P_K(k),~~~\forall k\in\calK .
\end{equation}
Suppose that we have already randomly selected a codebook for
one {\it representative} member $\hat{k}^n$ of each
type class $T_{Q_K}^0\subset
T_K^\delta$ 
using the mechanism described above.
Now, consider the
set of all permutations from $\hat{k}^n$ to every other member of 
$T_{Q_K}^0$.
The auxiliary codebook and the stegotext 
codebooks for every other key sequence, $k^n\in T_{Q_K}^0$
will be obtained by permuting all (auxiliary and stegotext) codewords of those
corresponding to $\hat{k}^n$ according
to the same permutation that leads from $\hat{k}^n$ to $k^n$ (thus preserving
all the necessary joint typicality properties).
Now, in the {\it union} of all stegotext codebooks, 
corresponding to all typical key
sequences, each codeword will appear 
at least $(n+1)^{-|\calK|\cdot|\calY|}
\cdot 2^{n[(1-\delta)H(K|Y)-\delta]}$ times,
which is a lower bound to the number of permutations of 
$\hat{k}^n$ which leave a
given stegotext codeword $y^n$ unaltered. 
The total number of stegotext codewords, $N_Y$,
in all codebooks of all $\delta$--typical key sequences (including
repetitions) is upper bounded by
\begin{eqnarray}
\label{Nyu}
N_Y&=&|T_K^\delta|\cdot N_c\nonumber\\
&\le&2^{n[(1+\delta)H(K)+\delta]}\cdot 
2^{n[\lambda R_U(D')+I(X;Y,V|K)+\epsilon_1+\epsilon_2+2\delta]}\nonumber\\
&=&2^{n[H(K)+
\lambda R_U(D')+I(X;Y,V|K)+\epsilon_1+\epsilon_2+\delta(H(K)+3)]}.
\end{eqnarray}
Let $\calC$ denote the 
union of all stegotext codebooks, namely, the set of all
{\it distinct} stegotext vectors
across all codebooks corresponding to all $k^n\in T_K^\delta$, and
let $N(y^n)$ denote the number of occurrences of a given vector 
$y^n\in\calY^n$ in all stegotext codebooks. Then,
in view of the above combinatorial consideration, we have
\begin{equation}
\label{Nyl}
N_Y=\sum_{y^n\in\calC} N(y^n)\ge |\calC|\cdot
(n+1)^{-|\calK|\cdot|\calY|}\cdot
2^{n[(1-\delta)H(K|Y)-\delta]}.
\end{equation}
Combining eqs.\ (\ref{Nyu}) and (\ref{Nyl}), we have
\begin{equation}
\label{66}
\log|\calC|\le n[\lambda R_U(D')+I(X;Y,V|K)+I(K;Y)+\delta'],
\end{equation}
where
\begin{equation}
\delta'=\epsilon_1+\epsilon_2+\delta(H(K)+H(K|Y)+4)+
|\calK|\cdot|\calY|\cdot\frac{\log(n+1)}{n},
\end{equation}
which is arbitrarily small provided that
$\delta$ is sufficiently small and $n$ is sufficiently large.
Thus, the rate required for 
public compression of $Y^n$ (without the key),
which is $(\log|\calC|)/n$,
is arbitrarily close to
$[\lambda R_U(D_1)+I(X;Y,V|K)+I(K;Y)]$, 
which in turn is upper bounded by $R_c$,
by condition (d) of Theorem 4.

Before we proceed to evaluate 
the equivocation levels,
an important comment is in order in the 
context of public compression (and a similar
comment will apply to private compression): 
Note that a straightforward 
(and not necessary optimal) method for public compression
of $Y^n$ is simply according to its 
index within $T_Y^\delta$, which requires about
$nH(Y)$ bits. On the other hand, 
the converse theorem tells us that the compressed
representation of $Y^n$ cannot be much shorter 
than $n[\lambda R_U(D')+I(X;Y,V|K)+I(K;Y)]$ bits
(cf.\ the necessity of condition 
(d) of Theorem 4). Thus, contradiction 
between these two facts is avoided
only if 
\begin{equation}
\label{inherent1}
\lambda R_U(D')+I(X;Y,V|K)+I(K;Y)\le H(Y),
\end{equation}
or, equivalently,
\begin{equation}
\label{inherent2}
\lambda R_U(D')+I(X;Y,V|K)\le H(Y|K).
\end{equation}
This means that any achievable point $(D,D',R_c,R_c',h,h')$
corresponds to a choice of random variables $(K,X,Y,V)$ that must 
inherently satisfy eq.\ (\ref{inherent2}).
This observation will now help us also in estimating the equivocation levels.

Consider first the equivocation w.r.t.\ the reproduction, 
for which we have the
following chain of inequalities:
\begin{eqnarray}
Nh'&\le&nH(K|Y)\label{cc}\\
&=&nH(K)-nI(K;Y)\nonumber\\
&=&H(K^n)-nI(K;Y)\label{dd}\\
&=&H(K^n|Y^n,Z^n)+I(K^n;Y^n,Z^n)-nI(K;Y)\nonumber\\
&=&H(K^n|Y^n,Z^n)+I(K^n;Y^n)-nI(K;Y)\label{aa}\\
&=&H(K^n|Y^n,Z^n)+H(Y^n)-H(Y^n|K^n)-nI(K;Y)\nonumber\\
&\le&H(K^n|Y^n,Z^n)+n[\lambda R_U(D')+I(X;Y,V|K)+I(K;Y)+\epsilon]-\nonumber\\
& &-n[\lambda R_U(D'+\epsilon)+I(X;Y,V|K)-\epsilon]-nI(K;Y)\label{bb}\\
&=&H(K^n|Y^n,Z^n)+n\lambda[R_U(D')-R_U(D'+\epsilon)]+n\epsilon\nonumber\\
&\dfn&H(K^n|Y^n,Z^n)+n\epsilon'\nonumber\\
&=&I(K^n;\hU^N|Y^n,Z^n)+H(K^n|Y^n,Z^n,\hU^N)+n\epsilon'\nonumber\\
&\le&H(\hU^N|Y^n,Z^n)+H(K^n|Y^n,Z^n,\hU^N)+n\epsilon'\label{last}
\end{eqnarray}
where (\ref{cc}) is based on condition (b),
(\ref{dd}) is due to the memorylessness of $K^n$,
(\ref{aa}) follows from the fact that $K^n\to Y^n\to Z^n$ is a Markov
chain, (\ref{bb}) is due to the 
sufficiency of condition (d) (that we have just proved)
and the necessity of condition (e), 
and $\epsilon'$ vanishes as $\epsilon\to 0$ due to the
continuity of $R_U(\cdot)$. 
Comparing the left--most side and the right--most side of the
above chain of inequalities, we see that 
to prove that $H(\hU^N|Y^n,Z^n)$ is essentially
at least as large as $Nh'$, it remains to show 
that $H(K^n|Y^n,Z^n,\hU^N)$ is small, say, 
\begin{equation}
\label{small}
H(K^n|Y^n,Z^n,\hU^N)\le
n\epsilon' 
\end{equation}
for large $n$. We next focus then on the proof of eq.\ (\ref{small}).

First, consider the following chain of inequalities:
\begin{eqnarray}
\label{equivbound}
H(K^n|Y^n,Z^n,\hat{U}^N)&\le& 
H(K^n,S^J(K^n)|Y^n,Z^n,\hat{U}^N)\nonumber\\
&=&H(S^J(K^n)|Y^n,Z^n,\hat{U}^N)+
H(K^n|S^J(K^n),Y^n,Z^n,\hat{U}^N)\nonumber\\
&\le&H(S^J(K^n)|Y^n,\hat{U}^N,W^L)+
H(K^n|S^J(K^n),Y^n),
\end{eqnarray}
where the second inequality follows from the fact that $W^L$ is function of $\hat{U}^N$
and the fact that conditioning reduces entropy.
As for the second term of the right--most side, we have by Fano's inequality
\begin{equation}
H(K^n|S^J(K^n),Y^n)\le 1+P_{\mbox{err}}\cdot n\log|\calK|\le 
n\epsilon'/2 ~~~\mbox{for large enough $n$},
\end{equation}
as $P_{\mbox{err}}\to 0$ is the probability of error associated with the
Slepian--Wolf decoder that estimates $K^n$ from its compressed version, $S^J(K^n)$, and the
``side information,'' $Y^n$. 
As for the first term of the right--most side of (\ref{equivbound}), we have
\begin{eqnarray}
H(S^J(K^n)|Y^n,\hat{U}^N,W^L)&=&H(W^L\oplus\tilde{W}^L|Y^n,\hat{U}^N,W^L)\nonumber\\
&\le& H(\tilde{W}^L|Y^n).
\end{eqnarray}
It remains to show that $H(\tilde{W}^L|Y^n)\le n\epsilon'/2$ as well. In order to show this, we
have to demonstrate that for a good code, once $Y^n$ is given, there is very little uncertainty
with regard to $\tilde{W}^L$, which is the index of the bin.

To this end, let us
suppose that the inequality in 
(\ref{inherent2}) is strict (otherwise, we can slightly increase
the allowable distortion level $D'$ and thus reduce $R_U(D')$).
As we prove in the Appendix, for any given (arbitrarily small) $\gamma > 0$, 
\begin{equation}
\label{doubleexp}
\mbox{Pr}\{\exists~y^n~\mbox{in the code of $\hat{k}^n$ that appears 
in more than $2^{n\gamma}$ bins}\}
\le |\calY|^n2^{-(n\gamma-\log e)2^{n\gamma}},
\end{equation}
that is, a double--exponential decay. The probability of the union of these events across all
representatives $\{\hat{k}^n\}$ of all $T_{Q_K}^0\subset T_K^\delta$ will just be multiplied by
the number of $\{T_{Q_K}^0\}$ in $T_K^\delta$, which is polynomial, and hence will continue to
decay double--exponentially.
Let us define then the event 
$$\{\exists~y^n~\mbox{in the stego--codebook of some $\hat{k}^n$ that appears in 
more than $2^{n\gamma}$ bins}\}$$
as yet another
error event (like the error events 1--5) that occurs with very small probability. Assume then, that
the randomly selected codebook is ``good'' in the sense that 
no stegovector appears in more than $2^{n\gamma}$ bins, for
any of the representatives $\{\hat{k}^n\}$. 
Now, given $y^n$, how many candidate bins
(corresponding to encrypted messages $\{\tilde{w}^L\}$) can be expected at most?
For a given $y^n$, let us confine attention to 
the $\delta$--conditional type class $T_{K|Y}^\delta(y^n)$ (key sequences outside this set
cannot have $y^n$ in their codebooks, as they are not jointly $\delta$--typical with $y^n$).
The conditional $\delta$--type class $T_{K|Y}^\delta(y^n)$ can be partitioned into  
conditional $0$--type classes $\{T_{Q_{K|Y}}^0(y^n)\}$, where $Q_{K|Y}$ exhausts the allowed
$\delta$--tolerance in the conditional distribution around $P_{K|Y}$, in the same spirit 
as before. Now, take an arbitrary representative $\tilde{k}^n$ from a given $T_{Q_{K|Y}}^0(y^n)$,
and consider the set of all permutations that lead from $\tilde{k}^n$ to all other
members $\{k^n\}$ of $T_{Q_{K|Y}}^0(y^n)$. Obviously, the stego--codebooks of all those
$\{k^n\}$ have exactly the same configuration 
of occurrences of $y^n$ as that of $\tilde{k}^n$ (since these permutations leave $y^n$ unaltered),
therefore they belong to exactly the same bins as in the codebook of $\tilde{k}^n$, the
number of which is at most $2^{\gamma n}$, by the hypothesis that we are using a good code.
In other words, as $k^n$ scans $T_{Q_{K|Y}}^0(y^n)$, there will be no new bins that
contain $y^n$ relative to those that are already in the codebook of $\tilde{k}^n$.
New bins that contain $y^n$ can be seen then only by scanning 
the other conditional $0$--types $\{T_{Q_{K|Y}}^0(y^n)\}$
within $T_{K|Y}^\delta(y^n)$, but the number such conditional $0$--types does not exceed
the total number of conditional $0$--types, which is upper bounded, in turn, by
$(n+1)^{|\calK|\cdot|\calY|}$ \cite{CK81}. Thus, the totality of stego--codebooks, for all
relevant $\{k^n\}$ cannot give more than $(n+1)^{|\calK|\cdot|\calY|}\cdot 2^{n\gamma}$ 
distinct bins
altogether. In other words, for a good codebook:
\begin{equation}
\label{zzz}
H(\tilde{W}^L|Y^n)\le \log[(n+1)^{|\calK|\cdot|\calY|}\cdot 2^{n\gamma}]=
n\left[\gamma+|\calK|\cdot|\calY|\cdot\frac{\log(n+1)}{n}\right]
\end{equation}
which is less than $n\epsilon'/2$ for an appropriate choice of $\gamma$ and for large enough $n$.

Finally, for the equivocation w.r.t.\ 
the original message source, we have 
the following:
\begin{eqnarray}
H(U^N|Y^n,Z^n)&=& 
H(\hat{U}^N|Y^n,Z^n)+H(U^N|Y^n,Z^n)-H(\hat{U}^N|Y^n,Z^n)\nonumber\\
&\ge&nH(K|Y)-2n\epsilon'+H(U^N|Y^n,Z^n)-H(\hat{U}^N|Y^n,Z^n)\nonumber\\
&=&nH(K|Y)+H(U^N)-I(U^N;\hat{U}^N)-I(U^N;Y^n,Z^n)-\nonumber\\
& &H(\hat{U}^N|U^N)+I(\hat{U}^N;Y^n,Z^n)-2n\epsilon'\nonumber\\
&\ge&nH(K|Y)+H(U^N)-H(\hat{U}^N)-I(U^N;Y^n,Z^n)-\nonumber\\
& &H(\hat{U}^N|U^N)+I(\hat{U}^N;Y^n,Z^n)-2n\epsilon'\nonumber\\
&\ge&nH(K|Y)+NH(U)-NR_U(D')-2\epsilon']-\nonumber\\
&&[I(U^N;Y^n,Z^n)+
H(\hat{U}^N|U^N)-I(\hat{U}^N;Y^n,Z^n)],
\end{eqnarray}
where first inequality is due to the fact that
$H(\hU^N|Y^n,Z^n)\ge n[H(K|Y)-2\epsilon']$, that we have just shown, 
and the third
is due to the memorylessness of $\{U_i\}$ and the fact
that the rate--distortion codebook 
size is $2^{NR_U(D')}$ and so, $H(\hat{U}^N)\le NR_U(D')$.
Now, the second bracketed expression on the right--most side is the
same as in eq.\ (\ref{bra}), where in the case of this specific scheme,
both inequalities in (\ref{bra}) become equalities, i.e., this expression
vanishes. This is because in our scheme, $U^N\to\hat{U}^N\to (Y^n,Z^n)$ 
is a Markov chain (and so, the first inequality of (\ref{bra}) is tight) and
because $H(\hat{U}^N|U^N,Y^n,Z^n)\le 
H(\hat{U}^N|U^N)=0$ (as $\hat{U}^N$ is a deterministic
function of $U^N$), which makes the second inequality of (\ref{bra}) tight.
As a result, we have
\begin{eqnarray}
H(U^N|Y^n,Z^n)&\ge&N[H(K|Y)/\lambda+H(U)-R_U(D')-2\epsilon'/\lambda]\nonumber\\
&\ge&N[h+R_U(D')-H(U)+H(U)-R_U(D')-2\epsilon/\lambda]\nonumber\\
&=&N(h-2\epsilon'/\lambda),
\end{eqnarray}
where we have used condition (a).
This completes the proof of the direct part.

\section*{Acknowledgements}
The author would like to thank Dr.\ Yossi
Steinberg for interesting discussions.
Useful comments made by the anonymous referees
are acknowledged with thanks.

\section*{Appendix}
\renewcommand{\theequation}{A.\arabic{equation}}
    \setcounter{equation}{0}

\noindent
{\it Proof of eq.\ (\ref{doubleexp})}.
The probability of obtaining $y^n$ in a single 
random selection within the codebook of $\hat{k}^n$ is given by
\begin{eqnarray}
\label{ff}
\mbox{Pr}\{Y^n(j',V^n(m,j,\hat{k}^n),\hat{k}^n)=y^n\}&=&
\frac{|T_{V|KY}^\delta(k^n,y^n)|}{|T_{V|K}^\delta(k^n)|}\cdot
\frac{1}{|T_{Y|KV}^\delta(k^n,v^n)|}\label{tt}\\
&\le&\frac{2^{n(1+\delta)H(V|K,Y)}}{2^{n[(1-\delta)H(V|K)-\delta]}}
\cdot\frac{1}{2^{n[(1-\delta)H(Y|K,V)-\delta]}}\nonumber\\
&=& 2^{-n[H(Y|K)-\delta'']},
\end{eqnarray}
where the first factor in the right--hand side of (\ref{tt}) 
is the probability
of having a $V^n(m,j,\hat{k}^n)=v^n$ that is typical with 
$y^n$ and $\hat{k}^n$ (a necessary condition for this $v^n$
to generate the given $y^n$), the second factor
is the probability of
selecting a given $y^n$ in the random 
selection of the steogtext code, and where
\begin{equation}
\delta''=\delta[H(V|K,Y)+H(V|K)+H(Y|K,V)+2].
\end{equation}
It now follows that the probability $q$ for at least one
occurrence of $y^n$ among the stegowords corresponding to
a certain bin, in the codebook of $\hat{k}^n$,
is upper bounded (using the union bound) by
\begin{eqnarray}
q&\le&M_2\cdot M_3\cdot
2^{-n[H(Y|K)-\delta'']}\nonumber\\
&=&2^{-n[H(Y|K)-I(X;V|K)-I(X;Y|V,K)-
\delta''-2\delta-\epsilon_1-\epsilon_2]}\nonumber\\
&=&2^{-n[H(Y|K)-I(X;V,Y|K)-
\delta''-2\delta-\epsilon_1-\epsilon_2]}\nonumber\\
&\dfn&2^{-n[H(Y|K)-I(X;Y,V|K)-\delta_1]}.
\end{eqnarray}
We are interested to upper bound the probability that a given $y^n$
appears as a stegoword in more than $2^{n\gamma}$ bins 
in the codebook of $\hat{k}^n$, for a given
$\gamma > 0$. For $i=1,\ldots,M_U$, let $A_i\in\{0,1\}$ be the
indicator function of the event 
$$\{y^n \mbox{appears as a stegoword in bin no.}~i~\mbox{at least once}\}.$$
Then, clearly $\{A_i\}$ are i.i.d.\ with $\mbox{Pr}\{A_i=1\}=q$.
Therefore,
\begin{eqnarray}
\mbox{Pr}\left\{\sum_{i=1}^{M_U}A_i \ge 2^{n\gamma}\right\}
&\le& \exp_2\left\{-M_UD\left(\frac{2^{n\gamma}}{M_U}\|q\right)\right\}\nonumber\\
&=& \exp_2\left\{-M_UD\left(2^{-n[\lambda R_U(D')-\gamma]}\|q\right)\right\},
\end{eqnarray}
where for $\alpha,\beta\in[0,1]$, the function $D(\alpha\|\beta)$
designates the binary divergence
\begin{equation}
D(\alpha\|\beta)=\alpha\log\frac{\alpha}{\beta}+
(1-\alpha)\log\frac{1-\alpha}{1-\beta}.
\end{equation}
Now, referring to eq.\ (\ref{inherent2}), suppose that
\begin{equation}
H(Y|K)\ge\lambda R_U(D')+I(X;V,Y|K)+\delta_1+2\gamma.
\end{equation}
Then, clearly,
\begin{equation}
2^{-n[\lambda R_U(D')-\gamma]} > 2^{-n[H(Y|K)-I(X;Y,V|K)-\delta_1]} \ge q
\end{equation}
and so, $\mbox{Pr}\{\sum_{i=1}^{M_U}A_i \ge 2^{n\gamma}\}$ is further upper bounded by
\begin{equation}
\mbox{Pr}\left\{\sum_{i=1}^{M_U}A_i \ge 2^{n\gamma}\right\}\le 
\exp_2\left\{-M_UD\left(2^{-n[\lambda R_U(D')-\gamma]}\|2^{-n[H(Y|K)-I(X;Y,V|K)-\delta_1]}\right)\right\}.
\end{equation}
To further bound this expression from above, we have to get a lower bound to
an expression of the form $D(e^{-na}\|e^{-nb})$ for $0< a < b$. Applying the
inequality $\log(1+x)=-\log(1-\frac{x}{1+x})\ge \frac{x\log e}{1+x}$,  for $x > -1$, we have:
\begin{eqnarray}
D(2^{-na}\|2^{-nb})&=& 2^{-na}\log\frac{2^{-na}}{2^{-nb}}+
(1- 2^{-na})\log\frac{1-2^{-na}}{1-2^{-nb}}\nonumber\\
&=&n(b-a)2^{-na}+(1- 2^{-na})\log\left(1+\frac{2^{-nb}-2^{-na}}{1-2^{-nb}}\right)\nonumber\\
&\ge&n(b-a)2^{-na}+(2^{-nb}-2^{-na})\log e\nonumber\\
&\ge&[n(b-a)-\log e]2^{-na}.
\end{eqnarray}
Applying this inequality with $a=\lambda R_U(D')-\gamma$ and $b=H(Y|K)-I(X;Y,V|K)-\delta_1$,
we get
\begin{equation}
D\left(2^{-n[\lambda R_U(D')-\gamma]}\|2^{-n[H(Y|K)-I(X;Y,V|K)-\delta_1]}\right)\ge
(n\gamma-\log e)2^{-n[\lambda R_U(D')-\gamma]}
\end{equation}
and so,
\begin{equation}
\mbox{Pr}\left\{\sum_{i=1}^{M_U}A_i \ge 2^{n\gamma}\right\}\le 2^{-(n\gamma-\log e)2^{n\gamma}},
\end{equation}
which decays double--exponentially rapidly with $n$. While, this inequality holds for
a {\it given} $y^n$, the probability that $\sum_{i=1}^{M_U}A_i \ge 2^{n\gamma}$ for {\it some}
$y^n\in\calY^n$ would be upper bounded, using the union bound, by
$|\calY|^n\cdot 2^{-(n\gamma-\log e)2^{n\gamma}}$, which still decays double--exponentially.
Thus, with very high probability the random selection of stegovectors, for $\hat{k}^n$,
is such that no stego codevector $y^n$ appears in more than $2^{n\gamma}$ bins.

\newpage

\end{document}

%% file: p90fig1.pstex_t
\begin{picture}(0,0)%
\includegraphics{p90fig1.pstex}%
\end{picture}%
\setlength{\unitlength}{3947sp}%
\begingroup\makeatletter\ifx\SetFigFont\undefined%
\gdef\SetFigFont#1#2#3#4#5{%
  \reset@font\fontsize{#1}{#2pt}%
  \fontfamily{#3}\fontseries{#4}\fontshape{#5}%
  \selectfont}%
\fi\endgroup%
\begin{picture}(5604,2694)(1426,-2874)
\put(1426,-1220){\makebox(0,0)[lb]{\smash{\SetFigFont{9}{10.8}{\rmdefault}{\mddefault}{\itdefault}{$U^N$}%
}}}
\put(2309,-300){\makebox(0,0)[lb]{\smash{\SetFigFont{9}{10.8}{\rmdefault}{\mddefault}{\itdefault}{$X^n$}%
}}}
\put(2972,-1220){\makebox(0,0)[lb]{\smash{\SetFigFont{9}{10.8}{\rmdefault}{\mddefault}{\itdefault}{$Y^n$}%
}}}
\put(2199,-2874){\makebox(0,0)[lb]{\smash{\SetFigFont{9}{10.8}{\rmdefault}{\mddefault}{\itdefault}{$K^n$}%
}}}
\put(4884,-1220){\makebox(0,0)[lb]{\smash{\SetFigFont{9}{10.8}{\rmdefault}{\mddefault}{\itdefault}{$Z^n$}%
}}}
\put(6540,-1182){\makebox(0,0)[lb]{\smash{\SetFigFont{9}{10.8}{\rmdefault}{\mddefault}{\itdefault}{$\hat{U}^N$}%
}}}
\end{picture}

%% file: p90fig2.pstex_t
\begin{picture}(0,0)%
\includegraphics{p90fig2.pstex}%
\end{picture}%
\setlength{\unitlength}{3947sp}%
\begingroup\makeatletter\ifx\SetFigFont\undefined%
\gdef\SetFigFont#1#2#3#4#5{%
  \reset@font\fontsize{#1}{#2pt}%
  \fontfamily{#3}\fontseries{#4}\fontshape{#5}%
  \selectfont}%
\fi\endgroup%
\begin{picture}(8187,4227)(151,-5398)
\put(151,-2180){\makebox(0,0)[lb]{\smash{\SetFigFont{12}{14.4}{\rmdefault}{\mddefault}{\itdefault}{$U^N$}%
}}}
\put(1807,-2148){\makebox(0,0)[lb]{\smash{\SetFigFont{12}{14.4}{\rmdefault}{\mddefault}{\itdefault}{$W^L$}%
}}}
\put(2555,-2148){\makebox(0,0)[lb]{\smash{\SetFigFont{12}{14.4}{\rmdefault}{\mddefault}{\itdefault}{$\tilde{W}^L$}%
}}}
\put(5815,-2148){\makebox(0,0)[lb]{\smash{\SetFigFont{12}{14.4}{\rmdefault}{\mddefault}{\itdefault}{$Z^n$}%
}}}
\put(6456,-3964){\makebox(0,0)[lb]{\smash{\SetFigFont{12}{14.4}{\rmdefault}{\mddefault}{\itdefault}{$\tilde{W}^L$}%
}}}
\put(2449,-2949){\makebox(0,0)[lb]{\smash{\SetFigFont{12}{14.4}{\rmdefault}{\mddefault}{\itdefault}{$S^J$}%
}}}
\put(2342,-2361){\makebox(0,0)[lb]{\smash{\SetFigFont{12}{14.4}{\rmdefault}{\mddefault}{\itdefault}{$\oplus$}%
}}}
\put(8026,-4636){\makebox(0,0)[lb]{\smash{\SetFigFont{12}{14.4}{\rmdefault}{\mddefault}{\itdefault}{$\hat{U}^N$}%
}}}
\put(6226,-5086){\makebox(0,0)[lb]{\smash{\SetFigFont{12}{14.4}{\rmdefault}{\mddefault}{\itdefault}{$W^L$}%
}}}
\put(4426,-2161){\makebox(0,0)[lb]{\smash{\SetFigFont{12}{14.4}{\rmdefault}{\mddefault}{\itdefault}{$Y^n$}%
}}}
\put(151,-4036){\makebox(0,0)[lb]{\smash{\SetFigFont{12}{14.4}{\rmdefault}{\mddefault}{\itdefault}{$K^n$}%
}}}
\put(3601,-1336){\makebox(0,0)[lb]{\smash{\SetFigFont{12}{14.4}{\rmdefault}{\mddefault}{\itdefault}{$X^n$}%
}}}
\put(5851,-4111){\makebox(0,0)[lb]{\smash{\SetFigFont{12}{14.4}{\rmdefault}{\mddefault}{\itdefault}{$S^J$}%
}}}
\put(6301,-4311){\makebox(0,0)[lb]{\smash{\SetFigFont{12}{14.4}{\rmdefault}{\mddefault}{\itdefault}{$\oplus$}%
}}}
\end{picture}